\documentclass[journal,10pt]{IEEEtran}

\usepackage{amssymb}
\usepackage{amsmath}
\usepackage{cite}
\usepackage{url}
\usepackage{xcolor}
\usepackage{cite,graphicx,amsmath,amssymb}
\usepackage{subfigure}
\usepackage{citesort}
\usepackage{fancyhdr}
\usepackage{mdwmath}
\usepackage{mdwtab}
\usepackage{caption}
\usepackage{amsthm}

\newtheorem{remark}{Remark}
\newtheorem{theorem}{Theorem}

\newtheorem{lemma}{Lemma}

\newtheorem{corollary}{Corollary}

\newtheorem{proposition}{Proposition}

\makeatletter
\def\ScaleIfNeeded{%
\ifdim\Gin@nat@width>\linewidth \linewidth \else \Gin@nat@width
\fi } \makeatother

\begin{document}

\title{Enhancing the Physical Layer Security of  Non-orthogonal Multiple Access in Large-Scale Networks}


\author{


Yuanwei~Liu,~\IEEEmembership{Student Member,~IEEE,}
        Zhijin~Qin,~\IEEEmembership{Student Member,~IEEE,}
        Maged~Elkashlan,~\IEEEmembership{Member,~IEEE,}
        Yue~Gao,~\IEEEmembership{Senior Member,~IEEE,}
        and Lajos~Hanzo,~\IEEEmembership{Fellow,~IEEE}
\thanks{Part of this work was presented at the IEEE International Conference on Communications (ICC), Kuala Lumpur, Malaysia, May 2016~\cite{qinphysical}.}
\thanks{Y. Liu, Z. Qin, M. Elkashlan, and Y. Gao are with Queen Mary University of London, London,
UK (email:\{yuanwei.liu, z.qin, maged.elkashlan, yue.gao\}@qmul.ac.uk).}
\thanks{L. Hanzo is with University of Southampton, Southampton,
UK (email:lh@ecs.soton.ac.uk).}
}
\maketitle
\begin{abstract}
This paper investigates the physical layer security of non-orthogonal multiple access (NOMA) in large-scale networks with invoking stochastic geometry. Both single-antenna and multiple-antenna aided transmission scenarios are considered, where the base station (BS) communicates with randomly distributed NOMA users. In the single-antenna scenario, we adopt a protected zone around the BS to establish an eavesdropper-exclusion area with the aid of careful channel-ordering of the NOMA users. In the multiple-antenna scenario, artificial noise is generated at the BS for further improving the security of a beamforming-aided system. In order to characterize the secrecy performance, we derive new exact expressions of the security outage probability  for both single-antenna and multiple-antenna aided scenarios. To obtain further insights, 1) for the single antenna scenario, we perform secrecy diversity order analysis of the selected user pair. The analytical results derived demonstrate that the secrecy diversity order is determined by the specific user having the worse channel condition among the selected user pair; and 2) for the multiple-antenna scenario, we derive the asymptotic secrecy outage probability, when the number of transmit antennas tends to infinity.  Monte Carlo simulations are provided for verifying the  analytical results derived and to show that: i)~The security performance of the NOMA networks can be improved by invoking the protected zone and by generating artificial noise at the BS; and ii)~The asymptotic secrecy outage probability is close to the exact secrecy outage probability.
\end{abstract}

\section{Introduction}
The unprecedented expansion of new Internet-enabled smart devices, applications and services is expediting the development of the  fifth generation (5G) networks, which aim for substantially increasing  the throughput of the fourth generation (4G) networks. In addition to the key technologies such as large-scale multiple-input multiple-output (MIMO) solutions, heterogeneous networks and millimeter wave, as well as novel multiple access (MA) techniques should be invoked for improving the spectral efficiency. The existing MA techniques can be primarily classified into two main categories, namely orthogonal multiple access and non-orthogonal multiple access (NOMA), by distinguishing whether a specific resource block can be occupied by more than one user~\cite{Wang2006OMA_NOMA}. More specifically, upon investigating the multiplexing gain gleaned from the different domains, the NOMA technique can be further classified as code-domain NOMA and power-domain NOMA~\cite{Dai2015NOMA}. Power-domain NOMA\footnote{Hence in this paper, we focus our attention on the family of power-domain NOMA schemes. We simply use ``NOMA" to refer to ``power-domain NOMA" in the following.}, which has been recently proposed for the 3GPP Long Term Evolution (LTE) initiative~\cite{3GPP}, is deemed to have a superior spectral efficiency~\cite{Saito:2013,Zhiguo2015Mag}. It has also been pointed out that NOMA has the potential to be integrated with existing MA paradigms, since it exploits the new dimension of the power domain. The key idea of NOMA is to ensure that multiple users can be served within a given resource slot (e.g., time/frequency), by applying successive interference cancellation (SIC). The concept of SIC, which was first proposed by Cover in 1972~\cite{cover1972broadcast}, constitutes  a promising technique, since it imposes lower complexity than the joint decoding approach~\cite{Andrews2005SIC}.

Hence NOMA techniques have received remarkable attention both in the world of academia and industry~\cite{ding2014performance,Timotheou:2015,yuanwei_JSAC_2015,Jinho:2015,Sunqi2015}.  Ding {\em et al.}~\cite{ding2014performance} investigated the performance of  the NOMA downlink for randomly roaming users. It was shown that NOMA is indeed capable of achieving a better performance than their traditional orthogonal multiple access (OMA) counter parts. By considering the user fairness of a NOMA system, a user-power allocation optimization problem was addressed by Timotheou and Krikidis~\cite{Timotheou:2015}. A cooperative simultaneous wireless power transfer (SWIPT) aided NOMA protocol was proposed by Liu {\em et al.}~\cite{yuanwei_JSAC_2015}, where a NOMA user benefitting from good channel conditions acts as an energy harvesting source in order to assist a NOMA user suffering from poor channel conditions. With the goal of maximizing the energy efficiency of transmission in multi-user downlink NOMA scenarios, Zhang {\em et al.}~\cite{Zhang2016NOMAEE} proposed an efficient power allocation technique capable of supporting the  data rate required by each user. To further improve the performance of NOMA systems, multiple antennas were introduced in~\cite{Jinho:2015,Sunqi2015}. More particularly, the application of multiple-input single-output (MISO) solution to NOMA was investigated by Choi {\em et al.}~\cite{Jinho:2015}, where a two-stage beamforming strategy was proposed.  Power optimization was invoked by Sun {\em et al.}~\cite{Sunqi2015} for maximizing the ergodic capacity of MIMO aided NOMA systems. As a further advance,  a  massive multiple-input multiple-output (MIMO) aided hybrid heterogenous  NOMA framework  was proposed for  downlink transmission by Liu {\em et al.}~\cite{yuanwei2016Hetnets}. The impact of the locations of users and interferers was investigated by using stochastic geometry approaches.


Given the broadcast nature of wireless transmissions, the concept of physical (PHY) layer security (PLS) was proposed by Wyner as early as 1975 from an information-theoretical perspective~\cite{wyner1975wire}. This research topic has sparked of wide-spread recent interests. To elaborate, PLS has been considered from  a practical perspective in~\cite{Mukherjee:2011,ZhiguoDing2012,Mukherjee:2011,Yuanwei:2015,zou2014security,liu2015secure}. Specifically, robust beamforming transmission was conceived in conjunction with applying artificial noise (AN) for mitigating the impact of imperfect channel state information (CSI) in MIMO wiretap channels was proposed by Mukherjee and Swindlehurst~\cite{Mukherjee:2011}. Ding~\emph{et al.}~\cite{ZhiguoDing2012} invoked relay-aided cooperative diversity for increasing the capacity of the desired link. More particularly, the impact of eavesdroppers on the diversity and multiplexing gains was investigated both in single-antenna and multiple-antenna scenarios.  Additionally, the tradeoffs between secure performance and reliability in the presence of eavesdropping attacks was identified by Zou~\emph{et al.}~\cite{zou2014security}. Furthermore, the physical layer security of D2D communication in large-scale cognitive radio networks was investigated by Liu~\emph{et al.}~\cite{liu2015secure} with invoking a wireless power transfer model,  where the positions of the power beacons, the legitimate and the eavesdropping nodes were modeled using stochastic geometry.

Recently, various PHY layer techniques, such as cooperative jamming~\cite{tekin2008general} and AN~\cite{goel2008guaranteeing} aided solutions were proposed for improving the PLS, even if the eavesdroppers have better channel conditions than the legitimate receivers. A popular technique is to generate AN at the transmitter for degrading the eavesdroppers' reception, which was proposed by Goel and Negi in~\cite{goel2008guaranteeing}. In contrast to the traditional view, which regards noise and interference as a detrimental effect,  generating AN at the transmitter is capable of improving the security, because it degrades the channel conditions of eavesdroppers without affecting those of the legitimate receivers. An AN-based multi-antenna aided secure transmission scheme affected by colluding eavesdroppers was considered by Zhou and McKay~\cite{zhou2010secure} for the scenarios associated both with perfect and imperfect CSI at both the transmitter and receiver. As a further development, the secrecy enhancement achieved in wireless Ad Hoc networks was investigated by Zhang {\em et al.}~\cite{zhang2013enhancing}, with the aid of both beamforming and sectoring techniques. By simultaneously considering matched filter precoding and AN generation techniques, the secure transmission strategies for a multi-user massive MIMO systems was investigated by Wu {\em et al.}~\cite{Yongpeng2016TIFS}. Very recently, the PLS of a single-input single-output (SISO) NOMA system was studied by Zhang {\em et al.}~\cite{Zhang2016PLS}, with the objective of maximizing the secrecy sum rate of multiple users.
\subsection{Motivation and Contribution}
As mentioned above,  PLS has been studied in various scenarios, but there is still a paucity of research contributions on investigating the security issues of NOMA, which motivates this contribution. Note that the employment of SIC in NOMA results in a unique interference status at the receivers, which makes the analysis of the PLS of NOMA  different from that of OMA. In this paper, we specifically consider the scenario of large-scale networks, where a base station (BS) supports  randomly roaming NOMA users. In order to avoid sophisticated high-complexity message detection at the receivers, a user pairing technique is adopted for ensuring that only two users share a specific orthogonal resource slot, which can be readily separated by low-complexity SIC. A random number of eavesdroppers are randomly positioned on an infinite two-dimensional plane according to a homogeneous Poisson point process (PPP).  An eavesdropper-exclusion zone is introduced around the BS for improving the secrecy performance of the large-scale networks considered in which no eavesdroppers are allowed to roam. This `disc' was referred to as a protected zone in~\cite{pinto2012secure,romero2013phy,zhang2013enhancing}. Specifically, we consider both a single-antenna scenario and a multiple-antenna scenario at the base station (BS). 1) For the single-antenna scenario, $M$ NOMA users are randomly roaming in an finite disc (user zone) with the quality-order of their channel conditions known at the BS. For example, the $m$-th NOMA user is channel-quality order of $m$. In this case, the $m$-th user is paired with the $n$-th user for transmission within the same resource slot; 2) For the multiple-antenna scenario, we invoke beamforming at the BS for generating AN. In order to reduce the complexity of channel ordering of MISO channels for NOMA, we partitioned the circular cell of Fig.~\ref{system_model} into an internal disc and an external ring. We select one user from the internal disc and another from the external ring to be paired together for transmission within the same resource slot using a NOMA protocol. The primary contributions of this paper are as follows:

\begin{itemize}
  \item  We investigate the secrecy performance of large-scale NOMA networks both for a single-antenna aided and a multiple-antenna assisted scenario at the BS. A protected zone synonymously referred to as the eavesdropper-exclusion area, is invoked in both scenarios for improving the PLS. Additionally, we propose to generate AN at the BS in the multiple-antenna aided scenario for further enhancing the secrecy performance.
  \item For the single-antenna scenario, we derive the exact analytical expressions of the secrecy outage probability~(SOP) of the selected pair of NOMA users, when relying on channel ordering. We then further extend on the secrecy diversity analysis and derive the expressions of asymptotic SOP. The  results  derived confirm that: 1) for the selected pair, the $m$-th user is capable of attaining a secrecy diversity order of $m$; 2) the secrecy diversity order is determined by the one associated  with the worse channel condition between the paired users.
  \item For the multiple-antenna scenario, we derive the exact analytical expressions of the SOP in conjunction with AN generated at the BS. To gain further insights, we assume having a large antenna array and derive the expressions of SOP, when the number of antennas tends to infinity. The results derived confirm that increasing the number of antennas has no effect on the received signal-to-interference-plus-noise ratio (SINR) at the eavesdroppers, when the BS is equipped with a large antenna array.
 \item  It is shown that: 1) the SOP can be reduced both by extending the protected zone and by generating AN at the BS; 2) the asymptotic SOP results of our large antenna array analysis is capable of closely approximating the exact secrecy outage probability; 3) there is an optimal desired signal-power and AN power sharing ratio, which minimizes the SOP in the multi-antenna scenario.
\end{itemize}
\subsection{Organization}
The rest of the paper is organized as follows. In Section II, a single-antenna transmission scenario is investigated in random wireless networks, where channel ordering of the NOMA users is relied on.  In Section III, a multiple-antenna transmission scenario is investigated, which relies on generating AN at the BS. Our numerical results are presented in Section IV for verifying our analysis, which is followed by our conclusions in Section V.

\section{Physical Layer Security in Random Wireless Networks with Channel Ordering}
\begin{figure} [t!]
\centering
\includegraphics[width= 2.3in, height=2.2in]{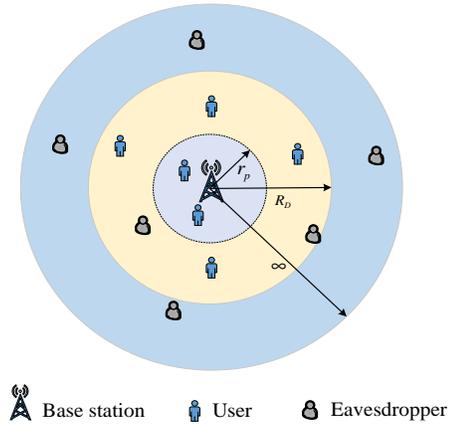}
 \caption{Network model for secure NOMA transmission in single-antenna scenario, where $r_p$, $R_D$, and $\infty$ is the radius of the Eve-exclusion area, NOMA user zone, and an infinite two dimensional plane for Eves, respectively.
  }
 \label{system_model}
\end{figure}

As shown in Fig.~\ref{system_model}, we focus our attention on a secure downlink communication scenario. In the scenario considered, a BS communicates with $M$ legitimate users (LUs) in the presence of eavesdroppers (Eves).  We assume that the $M$ users are divided into $M/2$ orthogonal pairs. Each pair is randomly allocated to a single resource block, such as a time slot or an orthogonal frequency band. For simplicity, we only focus our attention on investigating a typical pair of users in this treatise. Random user-pairing  is adopted in this work\footnote{We note that however sophisticated user pairing is capable of enhancing the performance of the  networks considered~\cite{yuanwei_JSAC_2015}, which is set aside for our future work.}. For each pair, the NOMA transmission protocol is invoked. It is assumed that BS is located at the center of a disc, denoted by $\mathcal{D}$, which has a coverage radius of ${R_{{D}}}$ (which is defined as the user zone for NOMA~\cite{ding2014performance}). The $M$ randomly roaming LUs are uniformly distributed within the disc. A random number of Eves is distributed across an infinite two-dimensional plane, which are assumed to have powerful detection capabilities and can overhear the messages of all orthogonal RBs, i.e. time slots or frequency slots. The spatial distribution of all Eves is modeled using a homogeneous PPP, which is denoted by ${\Phi_e}$ and it is associated with the density ${\lambda_e}$. It is assumed that the Eves can be detected, provided that they are close enough to BS. Therefore, an Eve-exclusion area having a radius of $r_p$ is introduced. Additionally, all channels are assumed to impose quasi-static Rayleigh fading, where the channel coefficients are constant for each transmission block, but vary independently between different blocks.

Without loss of generality, it is assumed that all the channels between the BS and LUs obey ${\left| {{h_1}} \right|^2} \le \cdot  \cdot  \cdot  {\left| {{h_m}} \right|^2} \le \cdot  \cdot  \cdot  {\left| {{h_n}} \right|^2} \le \cdot  \cdot  \cdot  {\left| {{h_M}} \right|^2}$. Both the the small-scale fading and the path loss are incorporated into the ordered channel gain. Again, we assume that the $m$-th user and the $n$-th user ($m < n$) are paired for transmission in the same resource slot.  Without loss of generality, we focus our attention on a single selected pair of users in the rest of the paper. In the NOMA transmission protocol, more power should be allocated to the user suffering from worse channel condition~\cite{Zhiguo2015Mag,Saito:2013}. Therefore, the power allocation coefficients satisfy the conditions that ${a_m} >  {a_n}$ and $a_m + a_n = 1$.  By stipulating this assumption, SIC can be invoked by the $n$-th user for first detecting the specific
  user having a higher transmit power (TX-power), who hence has a less
  interference-infested signal. Accordingly, the $m$-th user's signal
  is then remodulated and deducted from the original composite signal.
  This procedure then directly delivers the decontaminated
  lower-TX-power signal of the $n$-th user itself\footnote{It is assumed that perfect SIC is
  achieved at the $n$-th user, although achieving perfect SIC may be a non-trivial task. As a consequence, our analytical results may over-estimate the attainable secrecy performance of networks. Our future work will relax this
  idealized simplifying assumption, perhaps by analyzing both the connection outage probability and the secrecy outage probability of the networks considered, with the aid the  results derived in this treatise.}. We assume having fixed power allocation sharing between two users, but optimal power sharing strategies are capable of further enhancing the performance of the  networks considered, which is beyond the scope of this paper. Based on the aforementioned assumptions, the instantaneous SINR of the $m$-th user and the signal-to-noise ratio (SNR) of the $n$-th user can be written as:
\begin{align}\label{SINR B m}
{\gamma _{{B_m}}} = \frac{{{a_m}{{\left| {{h_m}} \right|}^2}}}{{{a_n}{{\left| {{h_m}} \right|}^2} + \frac{1}{\rho _b }}},
\end{align}
\begin{align}\label{SNR B n}
{\gamma _{{B_n}}} = \rho _b {a_n}{\left| {{h_n}} \right|^2},
\end{align}
respectively. We introduce the convenient concept of  transmit SNR ${\rho _b} = \frac{{{P_T}}}{{\sigma _b^2}}$, where $P_T$ is the TX-power of composite signal at the BS and  ${\sigma _b^2}$ is the variance of the additive white Gaussian noise (AWGN) at the LUs, noting that this is not a physically measurable quantity owing to their geographic separation. Perfectly flawless detection is assumed in this treatise, which is realistic at today's state-of-the-art with the aid of iterative turbo-detection techniques \cite{hanzo2010mimo,hanzo2011turbo}. Additionally, a bounded path loss model is used for guaranteeing that there is a practical path-loss, which is higher than one even for small distances.


We consider the worst-case scenario of large-scale networks, in which the Eves are assumed to have powerful detection capabilities~\cite{Koyluoglu2012PLS,zhang2013enhancing}. Specifically, by applying multiuser detection techniques, the multiuser data stream received from BS can be distinguished by the Eves, upon subtracting interference generated by the superposed signals from each other. In fact, this assumption overestimates the Eves' multi-user decodability.  In the scenario considered, all the CSIs of the LUs are assumed to be known at BS. However, the CSIs of Eves are assumed to be  unknown at the BS. The most detrimental Eve is not necessarily the nearest one, but the one having the best channel to BS. Non-colluding eavesdroppers are considered in this work. Therefore, the instantaneous SNR of detecting the information of the $m$-th user and the $n$-th user at the most detrimental Eve can be expressed as follows:
\begin{align}\label{SNR E P m_n}
{\gamma _{{E_\kappa }}} = \rho_e {a_\kappa }\mathop {\max }\limits_{e \in {\Phi _e},{d_e} \ge {r_p}} \left\{ {{{\left| {{g_e}} \right|}^2}L\left( {{d_e}} \right)} \right\}.
\end{align}
It is assumed that $\kappa  \in \left\{ {m,n} \right\}$, ${\rho _e} = \frac{{{P_T}}}{{\sigma _e^2}}$ is the transmit SNR with  ${\sigma _e^2}$ being the variance of the AWGN at Eves. Additionally, ${{{g}_e}}$ is defined as the small-scale fading coefficient associated with ${{{{g}_e}}} \sim \mathcal{CN}( {0,1})$, $L\left( {{d_e}} \right)= \frac{1}{{d_e ^\alpha }}$ is the path loss, and ${d_e }$ is the distance from Eves to BS. Note that due to the existence of the Eve-exclusion area (we assume $r_p>1$), it is not required to bound the path loss for Eves since ${d_e }$ will always be larger than one.
\subsection{New Channel Statistics}
In this subsection, we derive several new channel statistics for LUs and Eves, which will be used for deriving the secrecy outage probability in the next subsection.
\begin{lemma}\label{lemma:gamma Bn_CDF}
\emph{Assuming $M$ randomly located NOMA users in the disc of Fig.~\ref{system_model}, the cumulative distribution function (CDF) ${F_{{\gamma _{{B_n}}}}}$ of the $n$-th LU} \emph{is given by}
\begin{align}\label{gamma Bn_CDF}
&{F_{{\gamma _{{B_n}}}}}\left( x \right) \approx {{\varphi _n}}\sum\limits_{p = 0}^{M - n} {
 M - n \choose
 p } \frac{{{{\left( { - 1} \right)}^p}}}{{n + p}}\times\nonumber\\
 &\sum\limits_{\tilde S_n^p} {{
   {n + p}  \choose
   {{q_0} +  \cdots  + {q_K}}}\left( {\prod\limits_{K = 0}^K {b_k^{{q_k}}} } \right){e^{ - \sum\limits_{k = 0}^K {{q_k}} {c_k}\frac{x}{{\rho_b {a_n}}}}}},
\end{align}
\emph{where $K$ is a complexity-vs-accuracy tradeoff parameter}, ${b_k} =  - {\omega _K}\sqrt {1 - \phi _k^2} \left( {{\phi _k} + 1} \right)$, ${b_0} =  - \sum\limits_{k = 1}^K {{b_k}}$, ${c_k} = 1 + {\left[ {\frac{{{R_D}}}{2}\left( {{\phi _k} + 1} \right)} \right]^\alpha }$, ${\omega _K} = \frac{\pi }{K}$, ${\phi _k} = \cos \left( {\frac{{2k - 1}}{{2K}}\pi } \right)$, $\tilde S_n^p = \left\{ {\left. {\left( {{q_0},{q_1}, \cdots ,{q_K}} \right)} \right|\sum\limits_{i = 0}^K {{q_i}}  = n + p} \right\}$, ${ n + p \choose
 {q_0} +  \cdots  + {q_K} } = \frac{{\left( {n + p} \right)!}}{{{q_0}! \cdots {q_K}!}}$ \emph{and} ${\varphi _n} = \frac{{M!}}{{\left( {M - n} \right)!\left( {n - 1} \right)!}}$.
\begin{proof}
See Appendix A~.
\end{proof}
\end{lemma}

\begin{lemma}\label{lemma:gamma Bm CDF}
\emph{Assuming $M$ randomly positioned NOMA users in the disc of Fig.~\ref{system_model}, the CDF ${F_{{\gamma _{{B_m}}}}}$ of the $m$-th LU}  \emph{is given in~\eqref{gamma Bm CDF} at the top of next page,}
\begin{figure*}[!t]
\normalsize
\begin{align}\label{gamma Bm CDF}
{F_{{\gamma _{{B_m}}}}}\left( x \right)  \approx  U\left( {x - \frac{{{a_m}}}{{{a_n}}}} \right) + U\left( {\frac{{{a_m}}}{{{a_n}}} - x} \right){\varphi _m}\sum\limits_{p = 0}^{M - m} {
 M - m \choose
 p } \frac{{{{\left( { - 1} \right)}^p}}}{{m + p}}\sum\limits_{\tilde S_m^p} {{
 m + p \choose
 {q_0} +  \cdots  + {q_K}}\left( {\prod\limits_{k = 0}^K {b_k^{{q_k}}} } \right){e^{ - \sum\limits_{k = 0}^K {{q_k}{c_k}\frac{x}{{\left( {{a_m} - {a_n}x} \right)\rho_b }}} }}}.
\end{align}
\hrulefill \vspace*{0pt}
\end{figure*}
\emph{where $U\left( x \right) = \left\{ \begin{array}{l}
 1,x > 0 \\
 0,x \le 0 \\
 \end{array} \right.$ is the unit step function , and $\tilde S_m^p = \left\{ {\left. {\left( {{q_0},{q_1}, \cdots ,{q_K}} \right)} \right|\sum\limits_{i = 0}^K {{q_i}}  = m + p} \right\}$}.
\begin{proof}
Based on \eqref{SINR B m}, the CDF of ${F_{{\gamma _{{B_m}}}}}\left( x \right)$ can be expressed as
\begin{align}\label{gamma Bm_CDF proof 1}
{F_{{\gamma _{{B_m}}}}}\left( x \right) = \left\{ \begin{array}{l}
 \underbrace {\Pr \left\{ {{{\left| {{h_m}} \right|}^2} < \frac{x}{{\left( {{a_m} - {a_n}x} \right)\rho_b }}} \right\}}_{{\Phi _m}},x < \frac{{{a_m}}}{{{a_n}}} \\
 1,x \ge \frac{{{a_m}}}{{{a_n}}} \\
 \end{array} \right..
\end{align}
To derive the CDF of ${F_{{\gamma _{{B_m}}}}}\left( x \right)$, ${\Phi _m}$ can be expressed as ${\Phi _m} = {F_{{{\left| {{h_m}} \right|}^2}}}\left( {\frac{x}{{\left( {{a_m} - {a_n}x} \right)\rho_b }}} \right)$.
Based on \eqref{gamma Bn_ordered GC}, interchanging the parameters $m \to n$ and applying $y = \frac{x}{{\left( {{a_m} - {a_n}x} \right)\rho_b }}$, we obtain
\begin{align}\label{gamma Bm_Phi 1 2}
&{\Phi _m} = {\varphi _m}\sum\limits_{p = 0}^{M - m} {
 M - m \choose
 p } \frac{{{{\left( { - 1} \right)}^p}}}{{m + p}} \times\nonumber\\
& \sum\limits_{\tilde S_m^p}{
{ m + p \choose
 {q_0} +  \cdots  + {q_K}} \left( {\prod\limits_{k = 0}^K {b_k^{{q_k}}} } \right){e^{ - \sum\limits_{k = 0}^K {{q_k}{c_k}\frac{x}{{\left( {{a_m} - {a_n}x} \right)\rho_b }}} }}}.
\end{align}
By substituting \eqref{gamma Bm_Phi 1 2} into \eqref{gamma Bm_CDF proof 1}, with the aid of the unit step function, the CDF of ${F_{{\gamma _{{B_m}}}}}\left( x \right)$ can be obtained. The proof is completed.
\end{proof}
\end{lemma}

%
%

\begin{lemma}\label{lemma:gamma XE_PDF}
\emph{Assuming that the eavesdroppers obey the PPP distribution and the Eve-exclusion zone has a radius of $r_p$, the probability density function (PDF) ${f_{{\gamma _{{E_\kappa }}}}}$ of the most detrimental Eve} (where $\kappa  \in \left\{ {m,n} \right\}$
) \emph{is given by}
\begin{align}\label{gamma XE_PDF}
{f_{{\gamma _{{E_\kappa }}}}}\left( x \right) = {\mu _{\kappa 1}}{e^{ - \frac{{{\mu _{\kappa 1}}\Gamma \left( {\delta ,{\mu _{\kappa 2}}x} \right)}}{{{x^\delta }}}}}\left( {\frac{{\mu _{\kappa 2}^\delta {e^{ - {\mu _{\kappa 2}}x}}}}{x} + \frac{{\delta \Gamma \left( {\delta ,{\mu _{\kappa 2}}x} \right)}}{{{x^{\delta  + 1}}}}} \right),
\end{align}
\emph{where ${\mu _{\kappa 1}} = \delta \pi {\lambda _e}{\left( {{\rho _e}{a_\kappa }} \right)^\delta }, {\mu _{\kappa 2}} = \frac{{r_p^\alpha }}{{{\rho _e}{a_\kappa }}}$, $\delta  = \frac{2}{\alpha }$ and $\Gamma (\cdot,\cdot)$ is the upper incomplete Gamma function}.
\begin{proof}
To derive the PDF of ${f_{{\gamma _{{E_\kappa }}}}}\left( x \right)$, we have to compute the CDF of ${F_{{\gamma _{{E_\kappa }}}}}$ firstly as
\begin{align}\label{gamma XE_CDF 1}
{F_{{\gamma _{{E_\kappa }}}}}\left( x \right)
=& {\mathbb{E}_{{\Phi _e}}}\left\{ {\prod\limits_{e \in {\Phi _e},{d_e} \ge {r_p}} {{F_{{{\left| {{g_e}} \right|}^2}}}\left( {\frac{{xd_e^\alpha }}{{\rho_e {a_\kappa }}}} \right)} } \right\}.
\end{align}
Following the similar approach as \cite{Qin2017Tcom}, by applying the generating function~\cite{Stoyan}, \eqref{gamma XE_CDF 1} can be rewritten as
\begin{align}\label{gamma XE_CDF 2}
{F_{{\gamma _{{E_\kappa }}}}}\left( x \right) =& \exp \left[ { - {\lambda _e}{\int _{{R^2}}}\left( {1 - {F_{{{\left| {{g_e}} \right|}^2}}}\left( {\frac{{xd_e^\alpha }}{{\rho_e {a_\kappa }}}} \right)} \right)rdr} \right] \nonumber\\
 =& \exp \left[ { - 2\pi {\lambda _e}\int_{{r_p}}^\infty  {r{e^{ - \frac{x}{{\rho_e {a_\kappa }}}{r^\alpha }}}dr} } \right].
\end{align}
By applying \cite[ Eq. (3.381.9)]{gradshteyn}, we arrive at:
\begin{align}\label{gamma XE_CDF 3}
{F_{{\gamma _{{E_\kappa }}}}}\left( x \right) = {e^{ - \frac{{\delta \pi {\lambda _e}{{\left( {\rho_e {a_\kappa }} \right)}^\delta }\Gamma \left( {\delta ,\frac{{xr_p^\alpha }}{{\rho_e {a_\kappa }}}} \right)}}{{ {x^\delta }}}}}.
\end{align}

By taking the derivative of the CDF ${F_{{\gamma _{{E_\kappa }}}}}\left( x \right)$ in \eqref{gamma XE_CDF 3}, we obtain the PDF ${\gamma _{{E_\kappa }}}$ in \eqref{gamma XE_PDF}. The proof is completed.
\end{proof}
\end{lemma}

\subsection{Secrecy Outage Probability}

In the networks considered, the capacity of the LU's channel for the $\kappa $-h user ($\kappa  \in \left\{ {m,n} \right\}$
) is given by ${C_{{B_\kappa }}} = {\log _2}(1 + {\gamma _{{B_\kappa }}})$, while the capacity of the Eve's channel for the $\kappa $-th user is quantified by ${C_{{E_\kappa }}} = {\log _2}(1 + {\gamma _{{E_\kappa }}})$. It is assumed that the length of the block is sufficiently high for facilitating the employment of capacity-achieving codes within each block.  Additionally, the fading block length of the main channel and of the eavesdropper's channel are assumed to be the same. As such, according to \cite{Bloch2008IT}, the secrecy rate of the $n$-th and of the $m$-th user can be expressed as
\begin{align}\label{secrecy_rate n}
{C_n} = {\left[ {{C_{{B_n}}} - {C_{{E_n}}}} \right]^ + },
\end{align}
\begin{align}\label{secrecy_rate m}
{C_m} = {\left[ {{C_{{B_m}}} - {C_{{E_m}}}} \right]^ + }, 
\end{align}
where we have $[x]^+=\mathrm{max}\{x,0\}$. Here, the secrecy rates of LUs are strictly positive \cite{Jonas2013TIFS}. Recall that the Eves' CSIs are  unknown at the BS, hence the BS can only send information to the LUs at a constant rate, but perfect secrecy is not always guaranteed~\cite{wang2014physical}. Considering the $\kappa$-th user as an example, if ${R_\kappa} < {C_\kappa}$, the information with a rate of $R_\kappa$ ($\kappa  \in \left\{ {m,n} \right\}$ is conveyed in perfect secrecy. By contrast, for the case of ${R_\kappa} > {C_\kappa}$ the information-theoretic security is compromised. Motivated by this, secrecy outage probability is used as our secrecy performance metric in this paper. Given the expected secrecy rate ${ R}_{ {\kappa}}$ of the $\kappa$-th user, a secrecy outage event is declared, when the secrecy rate ${ C}_{ {\kappa}}$ drops below ${ R}_{ {\kappa}}$, which is defined as the SOP for the $\kappa$-th user. Recall that we have allocated $M$ users to $M/2$ orthogonal RBs, each pair of users are independent from all other pairs of users. We focus our attention on the SOP of a typical pair of users. We then derive the SOP of the $\kappa$-th user in the following two Theorems. We consider the SOP under the condition that the connection between BS and LUs can be established.

\begin{theorem}\label{SOP n_order}
\emph{Assuming that the LUs position obeys the PPP for the ordered channels of the LUs, the SOP of the $n$-th user is given by \eqref{Pout_n} at the top of this page.}
\begin{figure*}[!t]
\normalsize
\begin{align}\label{Pout_n}
P_n \left( {{R_n}} \right)=& {\varphi _n}\sum\limits_{p = 0}^{M - n} {M - n \choose
 p } \frac{{{{\left( { - 1} \right)}^p}}}{{n + p}}{\sum _{\tilde S_n^p}}{
 n + p \choose
 {q_0} +  \cdots  + {q_K} }\left( {\prod\limits_{K = 0}^K {b_k^{{q_k}}} } \right) \nonumber\\
& \times \int_0^\infty  {{\mu _{n1}}\left( {\frac{{\mu _{n 2}^\delta {e^{ - {\mu _{n 2}}x}}}}{x} + \frac{{\delta \Gamma \left( {\delta ,{\mu _{n 2}}x} \right)}}{{{x^{\delta  + 1}}}}} \right){e^{ - \frac{{{\mu _{n 1}}\Gamma \left( {\delta ,{\mu _{n 2}}x} \right)}}{{{x^\delta }}} - \sum\limits_{k = 0}^K {{q_k}} {c_k}\frac{{{2^{{R_n}}}\left( {1 + x} \right) - 1}}{{\rho_b {a_n}}}}}dx}.
\end{align}
\hrulefill \vspace*{0pt}
\end{figure*}
\begin{proof}
In this treatise, we consider the SOP under the condition that the connection between the BS and LUs can be established. As such, the SIC has been assumed to be successfully performed at the $n$-th user. Based on \eqref{secrecy_rate n}, the SOP is given by
\begin{align}\label{SOP n}
P_n\left( {{R_n}} \right) 
=& \int_0^\infty  {{f_{{\gamma _{{E_n}}}}}\left( x \right){F_{{\gamma _{{B_n}}}}}\left( {{2^{{R_n}}}\left( {1 + x} \right) - 1} \right)dx}.
\end{align}
Upon using the results of \textbf{Lemma \ref{lemma:gamma Bn_CDF}} and \textbf{Lemma~\ref{lemma:gamma XE_PDF}}, substituting \eqref{gamma Bn_CDF} and  \eqref{gamma XE_PDF} into \eqref{SOP n}, after some further mathematical manipulations, we can express the SOP of the $n$-th user. The proof is completed.
\end{proof}
\end{theorem}

\begin{theorem}\label{SOP m_order}
\emph{Assuming that the LUs position obeys the PPP for the ordered channels of the LUs, the SOP of the $m$-th user is given by \eqref{Pout_m} at the top of next page,}
\begin{figure*}[!t]
\normalsize
\begin{align}\label{Pout_m}
P_m \left( {{R_m}} \right)=& 1 - {e^{ - \frac{{{\mu _{m1}}\Gamma \left( {\delta ,{\tau _m}{\mu _{m2}}} \right)}}{{{\tau _m}^\delta }}}} + {\varphi _m}\sum\limits_{p = 0}^{M - m} {M - m \choose
 p } \frac{{{{\left( { - 1} \right)}^p}}}{{m + p}}{\sum _{\tilde S_m^p}}{
 m + p \choose
 {q_0} +  \cdots  + {q_K}}\left( {\prod\limits_{k = 0}^K {b_k^{{q_k}}} } \right) \nonumber\\
& \times \int_0^{{\tau _m}} {{\mu _{m1}}\left( {\frac{{\mu _{m2}^\delta {e^{ - {\mu _{m2}}x}}}}{x} + \frac{{\delta \Gamma \left( {\delta ,{\mu _{m2}}x} \right)}}{{{x^{\delta  + 1}}}}} \right){e^{ - \frac{{{\mu _{m1}}\Gamma \left( {\delta ,{\mu _{m2}}x} \right)}}{{{x^\delta }}} - \sum\limits_{k = 0}^K {{q_k}{c_k}\frac{{{2^{{R_m}}}\left( {1 + x} \right) - 1}}{{\left( {{a_m} - {a_n}\left( {{2^{{R_m}}}\left( {1 + x} \right) - 1} \right)} \right){\rho _b}}}} }}dx} .
\end{align}
\hrulefill \vspace*{0pt}
\end{figure*}
\emph{where we have ${\tau _m} = \frac{1}{{{2^{{R_m}}}\left( {1 - {a_m}} \right)}} - 1$.}
\begin{proof}
Based on \eqref{secrecy_rate m} and according to~\cite{Jonas2013TIFS}, the SOP for the $m$-th user is given by
\begin{align}\label{SOP m}
P_m\left( {{R_m}} \right) =& \int_0^\infty  {{f_{{\gamma _{{E_m}}}}}\left( x \right){F_{{\gamma _{{B_m}}}}}\left( {{2^{{R_m}}}\left( {1 + x} \right) - 1} \right)dx}.
\end{align}
Upon using the results of \textbf{Lemma \ref{lemma:gamma Bm CDF}} and \textbf{Lemma \ref{lemma:gamma XE_PDF}}, as well as substituting \eqref{gamma Bm CDF} and  \eqref{gamma XE_PDF} into \eqref{SOP m}, after some further mathematical manipulations, we can express the SOP of the $m$-th user. The proof is completed.
\end{proof}
\end{theorem}

In this paper, based on the assumptions of perfect SIC of LUs and strong detection capabilities of Eves aforementioned, the secrecy outage occurs in the $m$-th user and the $n$-th user are independent. Note that relaxing these two assumptions requires to consider dependence between two users by discussing more sophisticated connect/secrecy outage events, which should be included in our future work with the aid of the results derived in this paper. In other words, the SOP of the $m$-th user has no effect on the SOP of the $n$-th user and vice versa.  As a consequence, we define the SOP for the selected user pair as that of either the $m$-th user or the $n$-th user outage.
\begin{proposition}\label{proposition:SOP total}
The SOP of the selected user pair is given by
\begin{align}\label{Pout_total}
P_{mn} = 1 - \left( {1 - P_m} \right)\left( {1 - P_n} \right),
\end{align}
where $P_n$ and $P_m$ are given by \eqref{Pout_n} and \eqref{Pout_m}, respectively.
\end{proposition}


\subsection{Secrecy Diversity Order Analysis}
In order to derive the secrecy diversity order to gain further insights into the system's operation in the high-SNR regime, the following new analytical framework is introduced. Again, as the worst-case scenario, we assume that Eves have a powerful detection capability. The asymptotic behavior is analyzed, usually when the SNR of the channels between the BS and LUs is sufficiently high, i.e., when the BS's transmit SNR obeys ${\rho _b} \to \infty $, while and the SNR of the channels between BS and Eves is set to arbitrary values. It is noted that for the Eve's transmit SNR of ${\rho _e} \to \infty $, the probability of successful eavesdropping will tend to unity. The secrecy diversity order can be defined as follows:
\begin{align}\label{Diversity}
{d_s} = -\mathop {\lim }\limits_{{\rho _b} \to \infty } \frac{{\log P^\infty}}{{\log {\rho _b}}},
\end{align}
where $P^\infty$ is the asymptotic SOP.

\begin{corollary}\label{SOP n_asymptotic}
\emph{Assuming that the LUs position obeys the PPP for the ordered channels of the LUs, the asymptotic SOP of the $n$-th user is given by }
\begin{align}\label{SOP n asym}
P_n^\infty \left( {{R_n}} \right)
 &= {G_n}{\left( {{\rho _b}} \right)^{ - {D_n}}} + o\left( {\rho _b^{ - {D_n}}} \right),
\end{align}
\emph{where we have ${Q_1} = \int_0^\infty  {\mu _{n1}}{e^{ - \frac{{{\mu _{n1}}\Gamma \left( {\delta ,{\mu _{n2}}x} \right)}}{{{x^\delta }}}}}\times
\\\left( {\frac{{\mu _{n2}^\delta {e^{ - {\mu _{n2}}x}}}}{x} + \frac{{\delta \Gamma \left( {\delta ,{\mu _{n2}}x} \right)}}{{{x^{\delta  + 1}}}}} \right){\left( {\frac{{\left( {{2^{{R_n}}}\left( {1 + x} \right) - 1} \right)\ell }}{{{a_n}}}} \right)^{n}}dx$, ${G_n} = \frac{{{\varphi _n}{Q_1}}}{n}$, and ${D_n} = n$.}
\begin{proof}
We commence our diversity order analysis by characterizing the CDF of the LUs $F_{{\gamma _{{B_m}}}}^\infty $ and $F_{{\gamma _{{B_n}}}}^\infty $ in the high-SNR regime. When $y \to 0$, based on \eqref{gamma Bn_unordered 1} and the approximation of $1 - {e^{ - y}} \approx y$, we obtain the asymptotic unordered CDF of ${{{\left| {{{\tilde h}_n}} \right|}^2}}$ as follows:
\begin{align}\label{gamma Bn_unordered asym 1}
F_{{{\left| {{{\tilde h}_n}} \right|}^2}}^\infty \left( y \right) &\approx \frac{{2y}}{{R_D^2}}\int_0^{{R_D}} {\left( {1 + {r^\alpha }} \right)rdr} = y\ell,
\end{align}
where $\ell  = 1 + \frac{{2R_D^\alpha }}{{\alpha  + 2}}$.
Substituting \eqref{gamma Bn_unordered asym 1} into \eqref{gamma Bn_CDF2}, the asymptotic unordered CDF of ${{{\left| {{{\tilde h}_n}} \right|}^2}}$ is given by
\begin{align}\label{gamma Bn_ordered asym 1}
F_{{{\left| {{h_n}} \right|}^2}}^\infty \left( y \right) &= {\varphi _n}\sum\limits_{p = 0}^{M - n}
   {{M - n}  \choose
   p}  \frac{{{{\left( { - 1} \right)}^p}}}{{n + p}}{\left( {y\ell } \right)^{n + p}} \approx \frac{{{\varphi _n}}}{n}{\left( {y\ell } \right)^n}.
\end{align}
Then based on \eqref{gamma Bn_CDF1}, we can obtain $F_{{\gamma _{B_n}}}^\infty \left( x \right) \approx \frac{{{\varphi _n}}}{n}{\left( {\frac{{x\ell }}{{{\rho _b}{a_n}}}} \right)^n}$.
Based on \eqref{SOP n}, we can replace the CDF of ${F_{{\gamma _{{B_n}}}}}$ by the asymptotic $F_{{\gamma _{{B_n}}}}^\infty$. After some manipulations, we arrive at the asymptotic SOP of the $n$-th user.  The proof is completed.
\end{proof}
\begin{remark}\label{remark:diversity n}
Upon substituting \eqref{SOP n asym} into \eqref{Diversity}, we obtain the secrecy diversity order of the $n$-th user is $n$.
\end{remark}
\end{corollary}

\begin{corollary}\label{SOP m_asymptotic}
\emph{Assuming that the LUs position obeys the PPP for the ordered channels of the LUs, the asymptotic SOP for the $m$-th user is given by }
\begin{align}\label{SOP m asym}
P_m^\infty \left( {{R_n}} \right)&= {G_m}{\left( {{\rho _b}} \right)^{ - {D_m}}} + o\left( {\rho _b^{ - {D_m}}} \right),
\end{align}
\emph{where we have ${Q_2} = \int_0^{{\tau _m}}{{\mu _{m1}}{e^{ - \frac{{{\mu _{m1}}\Gamma \left( {\delta ,{\mu _{m2}}x} \right)}}{{{x^\delta }}}}}}\times
\\\left( {\frac{{\mu _{m2}^\delta {e^{ - {\mu _{m2}}x}}}}{x} + \frac{{\delta \Gamma \left( {\delta ,{\mu _{m2}}x} \right)}}{{{x^{\delta  + 1}}}}} \right){\left( {\frac{{\left( {{2^{{R_m}}}\left( {1 + x} \right) - 1} \right)\ell }}{{\left( {{a_m} - {a_n}\left( {{2^{{R_m}}}\left( {1 + x} \right) - 1} \right)} \right)}}} \right)^{m }}dx$, ${G_m} = \frac{{{\varphi _m}{Q_2}}}{m}$ and ${D_m} = m$.}
\begin{proof}
Based on ${\Phi _m}$ and \eqref{gamma Bn_ordered asym 1}, we can arrive at:
\begin{align}\label{Phi asym}
\Phi _m^\infty  \approx&\frac{{{\varphi _m}}}{m}{\left( {\frac{{x\ell }}{{\left( {{a_m} - {a_n}x} \right){\rho _b}}}} \right)^m}.
\end{align}
Substituting \eqref{Phi asym} into \eqref{gamma Bm_CDF proof 1}, the asymptotic CDF of ${{\gamma _{{B_m}}}}$ can be expressed as
\begin{align}\label{gamma Bm asym}
F_{{\gamma _{{B_m}}}}^\infty \left( x \right) = U\left( {x - \frac{{{a_m}}}{{{a_n}}}} \right) + U\left( {\frac{{{a_m}}}{{{a_n}}} - x} \right)\Phi _m^\infty,
\end{align}
where $\Phi _m^\infty$ is given in \eqref{Phi asym}. Then, based on \eqref{SOP m}, we can replace the CDF of ${F_{{\gamma _{{B_m}}}}}$ by the asymptotic $F_{{\gamma _{{B_m}}}}^\infty$ of \eqref{gamma Bm asym}. Additionally, we can formulate the asymptotic SOP of the $m$-th user. The proof is completed.
\end{proof}
\begin{remark}\label{remark:diversity m}
Upon substituting \eqref{SOP m asym} into \eqref{Diversity}, we obtain the secrecy diversity order of the $m$-th user is $m$.
\end{remark}
\end{corollary}

\begin{proposition}\label{proposition:diversity total}
\emph{For $m<n$, the secrecy diversity order can be expressed as}
\begin{align}\label{diversity total}
{d_s} =  - \mathop {\lim }\limits_{{\rho _b} \to \infty } \frac{{\log \left( {P_m^\infty  + P_n^\infty  - P_m^\infty P_n^\infty } \right)}}{{\log {\rho _b}}} = m.
\end{align}
\begin{proof}
Based on \textbf{Corollary~\ref{SOP m_asymptotic}} and \textbf{Corollary~\ref{SOP n_asymptotic}}, and upon substituting \eqref{SOP n asym} and \eqref{SOP m asym} into \eqref{Pout_total}, the asymptotic SOP for the user pair can be expressed as
\begin{align}\label{Pout_total asy}
P_{mn}^\infty  =& P_m^\infty  + P_n^\infty  - P_m^\infty P_n^\infty \approx P_m^\infty {G_m}{\left( {{\rho _b}} \right)^{ - {D_m}}}.
\end{align}
Upon substituting \eqref{Pout_total asy} into \eqref{Diversity}, we arrive at \eqref{diversity total}. The proof is completed.
\end{proof}
\end{proposition}
\begin{remark}\label{remark:diversity}
The results of \eqref{diversity total} indicate that the secrecy diversity order and the asymptotic SOP for the user pair considered are determined by the $m$-th user.
\end{remark}
\textbf{Remark~\ref{remark:diversity}} provides insightful guidelines for improving the SOP of the networks considered by invoking user pairing among of the $M$ users. Since the SOP of a user pair is determined by that of the one having a poor channel, it is efficient to pair the user having the best channel and the second best channel for the sake of achieving an increased secrecy diversity order.

\section{Enhancing Security with the aid of Artificial Noise}
\begin{figure} [t!]
\centering
\includegraphics[width= 2.6in, height=2.2in]{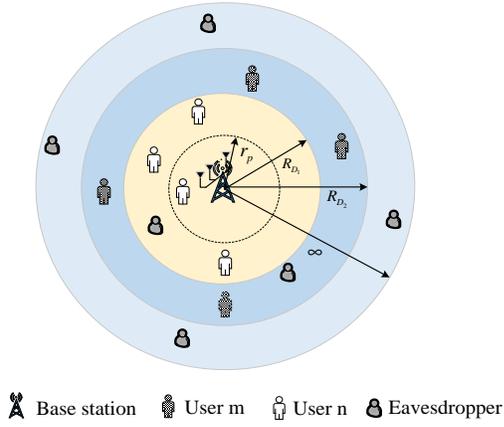}
 \caption{Network model for secure NOMA transmission using AN in multiple-antenna scenario, where $r_p$, $R_{D_1}$, $R_{D_2}$, and $\infty$ is the radius of the Eve-exclusion zone, NOMA user zone for user $n$, NOMA user zone for user $m$, and an infinite two dimensional plane for eavesdroppers, respectively.
  }
 \label{system_model_AN}
\end{figure}
In addition to single antenna scenario~\cite{qinphysical}, for further improving the secrecy performance, let us now consider the employment of  multiple antennas at BS for generating AN in order to degrade the Eves' SNR. More particularly, the BS is equipped with $N_A$ antennas, while all LUs and Eves are equipped with a single antenna each. Here, ${N_A} > 2$ is assumed for ensuring the existence of a null-space for two NOMA users. We mask the superposed information of NOMA by superimposing AN on Eves with the aid of the BS. It is assumed that the perfect CSI of LUs are known at BS\footnote{In practical scenarios, estimating the CSI may be a non-trivial task, therefore, our work actually provides an upper bound in terms of the attainable secrecy performance.}. Since the AN is in the null space of the intended LU's channel, it will not impose any effects on LUs. However, it can significantly degrade the channel and hence the capacity of Eves. More precisely, the key idea of using AN as proposed in~\cite{negi2008secret} can be described as follows: an orthogonal basis of $C^{N_A}$ is generated at BS for user $\kappa$, (where ${\kappa  \in \left\{ {m,n} \right\}}$) as a $\left( {{N_A} \times {N_A}} \right)$--element precoding matrix ${{\bf{U}}_{\kappa}} = \left[ {{{\bf{u}}_{\kappa}},{{\bf{V}}_{\kappa}}} \right]$, where we have ${{\bf{u}}_\kappa} = {\raise0.7ex\hbox{${\bf{h}_\kappa^\dag }$} \!\mathord{\left/
 {\vphantom {{\bf{h_\kappa}^\dag } {\left\| {{\bf{h}_\kappa}} \right\|}}}\right.\kern-\nulldelimiterspace}
\!\lower0.7ex\hbox{${\left\| {{\bf{h_\kappa}}} \right\|}$}}$ , and ${{\bf{V}}_{\kappa}}$ is  of size $N_A \times \left( {N_A - 1} \right)$. Here, ${\bf{h}_\kappa}$ is denoted as the intended channel between the BS and user $\kappa$. It is noted that each column of ${{\bf{V}}_{\kappa}}$ is orthogonal to ${{\bf{u}}_\kappa}$. Beamforming is applied at the BS for generating AN. As such, the transmitted superposed information, which is masked by AN at the BS is given by
\begin{align}\label{Signal_transmit}
\sum\limits_{\kappa  \in \left\{ {m,n} \right\}} {\sqrt {{a_\kappa }} {{\bf{x}}_\kappa }}  = \sum\limits_{\kappa  \in \left\{ {m,n} \right\}} \sqrt {{a_\kappa }} \left( {{s_\kappa }{{\bf{u}}_\kappa } + {{\bf{t}}_\kappa }{{\bf{V}}_\kappa }} \right),
\end{align}
where  $s_\kappa$ is the information-bearing signal with a variance of $\sigma _s^2$, and ${{\bf{t}}_\kappa}$ is the AN. Here the (${{N_A} - 1}$) elements of ${{\bf{t}}_\kappa}$ are independent identically distributed (i.i.d.) complex Gaussian random variables with a variance of $\sigma _a^2$. As such, the overall power per transmission is ${P_T} = {P_S} + {P_A}$, where ${P_S} = \theta {P_T} = \sigma _s^2$ is the transmission power of the desired information-bearing signal, while ${P_A} = \left( {1 - \theta } \right){P_T} = \left( {{N_A} - 1} \right)\sigma _a^2$ is the transmission power of the AN. Here $\theta$ represents the power sharing coefficients between the information-bearing signal and AN.

As shown in Fig.~\ref{system_model_AN}, we divide the disc $D$ into two regions, namely, $D_1$ and $D_2$, respectively. The motivation of using this topology hinges on two aspects. The first one is to create more distinct channel quality differences between the paired users, since existing NOMA studies have demonstrated that it is beneficial to pair two users having rather different channel conditions \cite{Saito:2013,ding2014pairing,yuanwei_JSAC_2015}. The second one is that of reducing the complexity of  channel ordering in this MISO NOMA system, which provides a compelling flexibility. By doing so, the path loss is the dominant channel impairment  in this scenario, because compared to the instantaneous small-scale fading effects, the path loss is more stable and more dominant. A quantitative example of comparing the small-scale fading and path loss was provided in Chapter 2 of \cite{steele1999mobile}. Note that the proposed design cannot guarantee the optimal ordering for MISO NOMA channels. More sophisticated precoding/detection design strategies (e.g., cluster based design, signal alignment and etc.) can be developed for further enhancing the attainable performance of the networks considered \cite{Yuanwei2016fairness,Zhiguo2016General}, but this is beyond the scope of this treatise. Here, $D_1$ is an internal disc with radius $R_{D_1}$, and the group of user $n$ is located in this region. $D_2$ is an external ring spanning the radius distance from $R_{D_1}$ to $R_{D_2}$, and the group of user $m$ is located in this region.  For simplicity, we assume that user $n$ and user $m$ are the selected user from each group in the rest of this paper. The cell-center user $n$ is assumed to be capable of cancelling the interference of the cell-edge user $m$ using SIC techniques\footnote{Note that upon invoking the signal alignment technique \cite{Zhiguo2016General}, the BS is capable of simultaneously supporting multiple pairs of NOMA users, by designing more sophisticated precoding/detection strategies for interference cancelation. However, these considerations are beyond the scope of this paper.}. User $n$ and user $m$ are randomly selected in each region for pairing them for NOMA. The combined signal at user $m$ is given by
\begin{align}\label{Receive signal m}
{{\bf{y}}_m} = \underbrace {\frac{{{{\bf{h}}_m}{{\bf{u}}_m}\sqrt {{a_m}} {s_m}}}{{\sqrt {1 + d_m^\alpha } }}}_{{\bf{Signal}}\;{\bf{part}}} + \underbrace {\frac{{{{\bf{h}}_m}{{\bf{u}}_n}\sqrt {{a_n}} {s_n}}}{{\sqrt {1 + d_m^\alpha } }} + \frac{{{{\bf{h}}_m}{{\bf{V}}_n}\sqrt {{a_n}} {{\bf{t}}_n}}}{{\sqrt {1 + d_m^\alpha } }} + {{\bf{n}}_m}}_{{\bf{Interference}}\;{\bf{and}}\;{\bf{noise}}\;{\bf{part}}},
\end{align}
where ${{\bf{n}}_m}$ is a Gaussian noise vector at user $m$, while $d_m$ is the distance between the BS and user $m$. Substituting \eqref{Signal_transmit} into \eqref{Receive signal m}, the received SINR at user $m$ is given by
\begin{align}\label{SINR m}
\gamma _{{B_m}}^{AN} = \frac{{{a_m}\sigma _s^2{{\left\| {{{\bf{h}}_m}} \right\|}^2}}}{{{a_n}\sigma _s^2{{\left| {{{\bf{h}}_m}\frac{{{\bf{h}}_n^\dag }}{{\left\| {{{\bf{h}}_n}} \right\|}}} \right|}^2} + {a_n}\sigma _a^2{{\left\| {{{\bf{h}}_m}{{\bf{V}}_n}} \right\|}^2} + 1 + d_m^\alpha }},
\end{align}
where the variance of ${{\bf{n}}_m}$ is normalized to unity. As such, we can express the transmit SNR at BS as $\rho_t=P_T$.

Since SIC is applied at user $n$, the interference arriving from user $m$ can be detected and subtracted firstly. The aggregate signal at user $n$ is given by
\begin{align}\label{Receive signal n}
{{\bf{y}}_n} = \underbrace {\frac{{{{\bf{h}}_n}{{\bf{u}}_n}\sqrt {{a_n}} {s_n}}}{{\sqrt {1 + d_n^\alpha } }}}_{{\bf{Signal}}\;{\bf{part}}} + \underbrace {\frac{{{{\bf{h}}_n}{{\bf{V}}_m}\sqrt {{a_m}} {{\bf{t}}_m}}}{{\sqrt {1 + d_n^\alpha } }} + {{\bf{n}}_n}}_{{\bf{Interference}}\;{\bf{and}}\;{\bf{noise}}\;{\bf{part}}},
\end{align}
where ${{\bf{n}}_n}$ is the Gaussian noise at user $n$, while $d_n$ is the distance between the BS and user $n$.  The received SINR at user $n$ is given by
\begin{align}\label{SINR n}
\gamma _{{B_n}}^{AN} = \frac{{{a_n}\sigma _s^2{{\left\| {{{\bf{h}}_n}} \right\|}^2}}}{{{a_m}\sigma _a^2{{\left\| {{{\bf{h}}_n}{{\bf{V}}_m}} \right\|}^2} + 1 + d_n^\alpha }},
\end{align}
where the variance of ${{\bf{n}}_n}$ is normalized to unity. The signal observed  by Eves is given by
\begin{align}\label{Receive signal eve}
{{\bf{y}}_e} = \sum\limits_{\kappa  \in \left\{ {m,n} \right\}} {\frac{{{{\bf{h}}_e}\sqrt {{a_\kappa }} {{\bf{x}}_\kappa }}}{{\sqrt {d_e^\alpha } }}}  + {{\bf{n}}_e},
\end{align}
where ${{\bf{n}}_e}$ is the Gaussian noises at Eves, while ${{\bf{h}}_e} \in {{\mathop{\rm C}\nolimits} ^{1 \times {N_A}}}$ is the channel vector between the BS and Eves. Similar to the single-antenna scenario, again, we assume that the Eves have a strong detection  capability and hence they unambiguously distinguish the messages of user $m$ and user $n$. The received SINR of the most detrimental Eve associated with detecting user $\kappa$ is given by
\begin{align}\label{SINR e}
\gamma _{{E_\kappa }}^{AN} = {a_\kappa }\sigma _s^2\mathop {\max }\limits_{e \in {\Phi _e},{d_e} \ge {r_p}} \left\{ {\frac{{{{X_{e,\kappa }}}}}{I_e^{AN} + d_e^\alpha }} \right\},
\end{align}
where the variance of ${{\bf{n}}_e}$ is normalized to unity, and we have ${X_{e,\kappa }} = {\left| {{{\bf{h}}_e}\frac{{{\bf{h}}_\kappa ^\dag }}{{\left\| {{{\bf{h}}_\kappa }} \right\|}}} \right|^2}$ as well as $I_e^{AN} = {a_m}\sigma _a^2{\left\| {{{\bf{h}}_e}{{\bf{V}}_m}} \right\|^2} + {a_n}\sigma _a^2{\left\| {{{\bf{h}}_e}{{\bf{V}}_n}} \right\|^2}$.

\subsection{New Channel Statistics}
In this subsection, we derive several new channel statistics for LUs and Eves in the presence of AN, which will be used for deriving the SOP in the next subsection.

\begin{lemma}\label{lemma:CDF Bn AN}
\emph{Assuming that user $n$ is randomly positioned in the disc $D_1$ of Fig.~\ref{system_model_AN}, the CDF of $F_{{B_n}}^{AN}$ is given by  }
\begin{align}\label{CDF Bn AN final}
&F_{{B_n}}^{AN}\left( x \right) = 1 - {b_2}{e^{ - \frac{{\vartheta x}}{{{a_m}}}}}\sum\limits_{p = 0}^{{N_A} - 1} {\frac{{{\vartheta ^p}{x^p}}}{{p!}}} \sum\limits_{q = 0}^p {
 p \choose q }\times\nonumber\\
& \frac{{\Gamma \left( {{N_A} - 1 + q} \right)a_m^{q - p}}}{{{{\left( {\vartheta x + \frac{{{N_A} - 1}}{{{P_A}}}} \right)}^{{N_A} - 1 + q}}}}\sum\limits_{u = 0}^{p - q} {
p - q \choose
u
} \frac{{a_m^{u + \delta }\gamma \left( {u + \delta ,\frac{{\vartheta x}}{{{a_m}}}R_{{D_1}}^\alpha } \right)}}{{{{\left( {\vartheta x} \right)}^{u + \delta }}}},
\end{align}
\emph{where we have ${b_2} = \frac{\delta }{{R_{{D_1}}^2\Gamma \left( {{N_A} - 1} \right){{\left( {\frac{{{P_A}}}{{{N_A} - 1}}} \right)}^{{N_A} - 1}}}}$ and $\vartheta  = \frac{{{a_m}}}{{{a_n}{P_S}}}$.}
\begin{proof}
See Appendix~B.
\end{proof}
\end{lemma}

\begin{lemma}\label{lemma:CDF Bm AN}
\emph{Assuming that user $m$ is randomly positioned in the ring $D_2$ of Fig.~\ref{system_model_AN}, for the case of ${\theta  \ne \frac{1}{{{N_A}}}}$, the CDF of $F_{{B_m}}^{AN}$ is given by}
\begin{align}\label{CDF Bm AN final}
&F_{{B_m}}^{AN}\left( x \right) = 1 - {e^{ - \frac{{\nu x}}{{{a_n}}}}}\sum\limits_{p = 0}^{{N_A} - 1} {\frac{{{{\left( {\nu x} \right)}^p}}}{{p!}}} \sum\limits_{q = 0}^p {
 p \choose
 q } a_n^{q - p} \times \nonumber\\
& \underbrace {{a_1}\left( {\frac{{\Gamma \left( {q + 1} \right)}}{{{{\left( {\nu x + \frac{1}{{{P_S}}}} \right)}^{q + 1}}}} - \sum\limits_{l = 0}^{{N_A} - 2} {\frac{{{{\left( {\frac{{{N_A} - 1}}{{{P_A}}} - \frac{1}{{{P_S}}}} \right)}^l}}}{{\frac{{l!{{\left( {\nu x + \frac{{{N_A} - 1}}{{{P_A}}}} \right)}^{q + l + 1}}}}{{\Gamma \left( {q + l + 1} \right)}}}}} } \right)}_{{\rm I}\left( \theta  \right)}\times\nonumber\\
&\sum\limits_{u = 0}^{p - q} {
 p - q \choose
 u } \frac{{\gamma \left( {u + \delta ,\frac{{\nu x}}{{{a_n}}}R_{{D_2}}^\alpha } \right) - \gamma \left( {u + \delta ,\frac{{\nu x}}{{{a_n}}}R_{{D_1}}^\alpha } \right)}}{{{{\left( {\frac{{\nu x}}{{{a_n}}}} \right)}^{u + \delta }}}},
\end{align}
\emph{where $\gamma \left( { \cdot , \cdot } \right)$ is the lower incomplete Gamma function, $\Gamma \left( {\cdot } \right)$
 is the Gamma function, ${a_1} = \delta {\left( {1 - \frac{{{P_A}}}{{\left( {{N_A} - 1} \right){P_S}}}} \right)^{1 - {N_A}}}/\left( {\left( {R_{{D_2}}^2 - R_{{D_1}}^2} \right){P_S}} \right)$,  and $\nu  = \frac{{{a_n}}}{{{a_m}{P_S}}}$.}

\emph{For the case of ${\theta  = \frac{1}{{{N_A}}}}$, the CDF of $F_{{B_m}}^{AN}$ is given by \eqref{CDF Bm AN final} upon substituting ${{\rm I}\left( \theta  \right)}$ by ${{\rm I}^*}\left( \theta  \right) $, where we have ${{\rm I}^*}\left( \theta  \right) = \frac{{{a_2}\Gamma \left( {q + {N_A}} \right)}}{{{{\left( {\nu x + \frac{1}{{{P_S}}}} \right)}^{q + {N_A}}}}}\sum\limits_{u = 0}^{p - q} {
 p - q \choose u }$ and ${a_2} = \frac{\delta }{{\left( {R_{{D_2}}^2 - R_{{D_1}}^2} \right){P_S}^{{N_A}}\left( {{N_A} - 1} \right)!}}$.}
\begin{proof}
See Appendix~C.
\end{proof}
\end{lemma}

\begin{lemma}\label{lemma:CDF Ek AN}
\emph{Assuming that the distribution of Eves obeys a PPP and that the Eve-exclusion zone has a radius of $r_p$, the PDF of ${f_{\gamma _{{E_\kappa }}^{AN}}}$ (where ${\kappa  \in \left\{ {m,n} \right\}}$) is given by}
\begin{align}\label{PDF Ek AN final}
&{f_{\gamma _{{E_\kappa}}^{AN}}}\left( x \right) =  - {e^{{\Theta _\kappa }{\Psi _{\kappa 1}}}} \times \nonumber\\
&\left( {\frac{{{{\left( {\mu _{\kappa 2}^{AN}} \right)}^\delta }{e^{ - x\mu _{\kappa 2}^{AN}}}}}{x}{\Psi _{\kappa 1}} + \frac{{\delta {\Theta _\kappa }{\Psi _{\kappa 1}}}}{x} + {\Theta _\kappa }{\Psi _{\kappa 2}}} \right),
\end{align}
\emph{where ${\Theta _\kappa } = \frac{{\Gamma \left( {\delta ,x\mu _{\kappa 2}^{AN}} \right)}}{{{x^\delta }}}$, $\Gamma \left( { \cdot , \cdot } \right)$ is the upper incomplete Gamma function, ${\Psi _{\kappa 1}} = \Omega \frac{1}{{{{\left( {\frac{x}{{{a_\kappa }{P_S}}} + {\tau _i}} \right)}^j}}},{\Psi _{\kappa 2}} = \Omega \frac{1}{{{{\left( {\frac{x}{{{a_\kappa }{P_S}}} + {\tau _i}} \right)}^j}}}\left( {\frac{j}{{\left( {\frac{x}{{{a_\kappa }{P_S}}} + {\tau _i}} \right)}}\frac{1}{{{a_\kappa }{P_S}}}} \right)$, $\Omega  = {\left( { - 1} \right)^{{N_A}}}\mu _{\kappa 1}^{AN}\times\\
\prod\limits_{i = 1}^2 {\tau _i^{{N_A} - 1}} \sum\limits_{i = 1}^2 {\sum\limits_{j = 1}^{{N_A} - 1} {{a_{{N_A} - j,{N_A} - 1}}{{\left( {2{\tau _i} - L} \right)}^{j - \left( {2{N_A} - 2} \right)}}} } $ , $L = {\tau _1} + {\tau _2},{\tau _1} = \frac{{{N_A} - 1}}{{{a_m}{P_A}}},{\tau _2} = \frac{{{N_A} - 1}}{{{a_n}{P_A}}}$, ${a_{{N_A} - j,{N_A} - 1}} ={
 2{N_A} - j - 3 \choose
 {N_A} - j - 1 }$, $\mu _{\kappa 1}^{AN} = \pi {\lambda _e}\delta {\left( {{a_\kappa }{P_S}} \right)^\delta }$, and $\mu _{\kappa 2}^{AN} = \frac{{r_p^\alpha }}{{{a_\kappa }{P_S}}}$.}

\begin{proof}
See Appendix~D.
\end{proof}
\end{lemma}


\subsection{Secrecy Outage Probability}
In this subsection, we investigate the SOP of a multiple-antenna aided scenario relying on AN.

\begin{theorem}\label{SOP AN n}
\emph{Assuming that the LUs and Eves distribution obey PPPs and that AN is generated at the BS, the SOP of user $n$ is given by \eqref{SOP n AN} at the top of next page,}
\begin{figure*}[!t]
\normalsize
\begin{align}\label{SOP n AN}
P_n^{AN}\left( {{R_n}} \right) &= \int_0^\infty  { - {e^{{\Theta _n}{\Psi _{n1}}}}\left( {\frac{{{{\left( {\mu _{n2}^{AN}} \right)}^\delta }{e^{ - x\mu _{n2}^{AN}}}}}{x}{\Psi _{n1}} + \frac{{\delta {\Theta _n}{\Psi _{n1}}}}{x} + {\Theta _n}{\Psi _{n2}}} \right)}  \nonumber\\
&\times\left( {1 - {b_2}{e^{ - {\iota _n}}}\sum\limits_{p = 0}^{{N_A} - 1} {\frac{{{\iota _{n*}}}}{{p!}}} \sum\limits_{q = 0}^p {
 p \choose
 q } \frac{{\Gamma \left( {{N_A} - 1 + q} \right)a_m^q}}{{{{\left( {{a_m}{\iota _{n*}} + \frac{{{N_A} - 1}}{{{P_A}}}} \right)}^{{N_A} - 1 + q}}}}\sum\limits_{u = 0}^{p - q} {
 p - q \choose
 u } \frac{{\gamma \left( {u + \delta ,{\iota _{n*}}R_{{D_1}}^\alpha } \right)}}{{\iota _{n*}^{u + \delta }}}} \right)dx,
\end{align}
\hrulefill \vspace*{0pt}
\end{figure*}
\emph{where ${\iota _{n*}} = \frac{{\vartheta \left( {{2^{{R_n}}}\left( {1 + x} \right) - 1} \right)}}{{{a_m}}}$}.
\begin{proof}
Using the results of \textbf{Lemma \ref{lemma:CDF Bn AN}} and \textbf{Lemma \ref{lemma:CDF Ek AN}},  upon substituting \eqref{CDF Bn AN final} and \eqref{PDF Ek AN final} into \eqref{SOP n}, we  can obtain the SOP of user $n$. The proof is completed.
\end{proof}
\end{theorem}

\begin{theorem}\label{SOP AN m}
\emph{Assuming that the LUs and Eves distribution obey PPPs and  that AN is generated at the BS, for the case $\theta  \ne \frac{1}{{{N_A}}}$, the SOP of user $m$ is given by \eqref{SOP m AN} at the top of next page,}
\begin{figure*}[!t]
\normalsize
\begin{align}\label{SOP m AN}
 P_m^{AN}\left( {{R_m}} \right) &= \int_0^\infty  { - {e^{{\Theta _m}{\Psi _{m1}}}}\left( {\frac{{{{\left( {\mu _{m2}^{AN}} \right)}^\delta }{e^{ - x\mu _{m2}^{AN}}}}}{x}{\Psi _{m1}} + \frac{{\delta {\Theta _m}{\Psi _{m1}}}}{x} + {\Theta _m}{\Psi _{m2}}} \right)}    \nonumber\\
&\times \underbrace {\left( {1 - a_1^*\sum\limits_{p = 0}^{{N_A} - 1} {\frac{{\iota _m^p}}{{p!}}} \sum\limits_{q = 0}^p {
 p \choose
 q } a_n^q\left( {\frac{{\Gamma \left( {q + 1} \right)}}{{{{\left( {{a_n}{\iota _{m*}} + \frac{1}{{{P_S}}}} \right)}^{q + 1}}}} - \sum\limits_{l = 0}^{{N_A} - 2} {\frac{{\frac{1}{{l!}}{{\left( {\frac{{{N_A} - 1}}{{{P_A}}} - \frac{1}{{{P_S}}}} \right)}^l}\Gamma \left( {q + l + 1} \right)}}{{{{\left( {{a_n}{\iota _{m*}} + \frac{{{N_A} - 1}}{{{P_A}}}} \right)}^{q + l + 1}}}}} } \right){\rm T}_1^*} \right)}_{{\rm K}\left( \theta  \right)}dx,
\end{align}
\hrulefill \vspace*{0pt}
\end{figure*}
\emph{where we have $a_1^* = \frac{{\delta {e^{ - {\iota _{m*}}}}{{\left( {1 - \frac{{{P_A}}}{{\left( {{N_A} - 1} \right){P_S}}}} \right)}^{1 - {N_A}}}}}{{\left( {R_{{D_2}}^2 - R_{{D_1}}^2} \right){P_S}}}$, ${\rm T}_1^* = \sum\limits_{u = 0}^{p - q} {
 p - q \choose
 u } \frac{{\gamma \left( {u + \delta ,{\iota _{m*}}R_{{D_2}}^\alpha } \right) - \gamma \left( {u + \delta ,{\iota _{m*}}R_{{D_1}}^\alpha } \right)}}{{\iota _{m*}^{u + \delta }}}$, and ${\iota _{m*}} = \frac{{\nu \left( {{2^{{R_m}}}\left( {1 + x} \right) - 1} \right)}}{{{a_n}}}$.}

\emph{For the case of ${\theta  = \frac{1}{{{N_A}}}}$, the SOP for user $m$ is given by \eqref{SOP m AN} upon substituting ${{\rm K}\left( \theta  \right)}$ with ${{\rm K}^*}\left( \theta  \right)$, where ${{\rm K}^*}\left( \theta  \right) = 1 - a_2^*\sum\limits_{p = 0}^{{N_A} - 1} {\frac{{\iota _{m*}^p}}{{p!}}} \sum\limits_{q = 0}^p {
 p \choose
 q } \frac{{\Gamma \left( {q + {N_A}} \right)a_n^q}}{{{{\left( {{a_n}{\iota _{m*}} + \frac{1}{{{P_S}}}} \right)}^{q + {N_A}}}}}\sum\limits_{u = 0}^{p - q} {
 p - q \choose
 u } {\rm T}_1^*$, and $a_2^* = \frac{{\delta {e^{ - {\iota _{m*}}}}}}{{\left( {R_{{D_2}}^2 - R_{{D_1}}^2} \right){P_S}^{{N_A}}\left( {{N_A} - 1} \right)!}}$}.
\begin{proof}
Using the results of \textbf{Lemma \ref{lemma:CDF Bm AN}} and \textbf{Lemma \ref{lemma:CDF Ek AN}}, upon substituting \eqref{CDF Bm AN final} and \eqref{PDF Ek AN final} into \eqref{SOP m}, we obtain the SOP of user $m$. The proof is completed.
\end{proof}
\end{theorem}

\begin{proposition}\label{proposition:SOP AN total}
The SOP of multiple-antenna aided scenario relaying on AN for the selected user pair can be expressed as
\begin{align}\label{Pout_total AN}
P^{AN}_{mn} = 1 - \left( {1 - P^{AN}_m} \right)\left( {1 - P^{AN}_n} \right).
\end{align}
where $P_n$ and $P_m$ are given by \eqref{SOP n AN} and \eqref{SOP m AN}, respectively.
\end{proposition}

\subsection{Large Antenna Array Analysis}
In this subsection, we investigate the system's asymptotic behavior when the BS is equipped with large antenna arrays. Large antenna arrays using narrow beamforming are potentially capable of  distinguishing multiple users in the angular domain~\cite{Xie2016JSAC,Xie2016access}. Nonetheless, the users covered by the same narrow beam in dense deployments still remain  non-orthogonal~\cite{Zhiguo2016massive}. It is noted that for the exact SOP derived in \eqref{SOP m AN} and \eqref{SOP n AN}, as $N_A$ increases, the number of summations in the equations will increase exponentially, which imposes an excessive complexity. Motivated by this, we seek good approximations for the SOP associated with a large $N_A$. With the aid of the theorem of large values, we have the following approximations~\cite{zhou2010secure}. $\mathop {\lim }\limits_{{N_A} \to \infty } {\left\| {{{\bf{h}}_n}} \right\|^2} \to {N_A}$, $\mathop {\lim }\limits_{{N_A} \to \infty } {\left\| {{{\bf{h}}_m}} \right\|^2} \to {N_A}$, $\mathop {\lim }\limits_{{N_A} \to \infty } {\left\| {{{\bf{h}}_n}{{\bf{V}}_m}} \right\|^2} \to {N_A} - 1$, and $\mathop {\lim }\limits_{{N_A} \to \infty } {\left\| {{{\bf{h}}_m}{{\bf{V}}_n}} \right\|^2} \to {N_A} - 1$.
We first derive the asymptotic CDF of user $n$ for ${N_A} \to \infty$.
\begin{lemma}\label{lemma:CDF Bn large final}
\emph{Assuming that user $n$ is randomly located in the disc $D_1$ of Fig.~\ref{system_model_AN} and ${N_A} \to \infty$, the CDF of $F_{{B_n},\infty }^{AN}$ is given by}
\begin{align}\label{CDF Bn large 2}
F_{{B_n},\infty }^{AN}\left( x \right) =  \left\{ \begin{array}{l}
 0,x < {\zeta _n} \\
 1 - \frac{{{{\left( {\frac{{{a_n}{P_S}{N_A}}}{x} - {a_m}{P_A} - 1} \right)}^\delta }}}{{R_{{D_1}}^2}},{\zeta _n} \le x \le {\xi _n} \\
 1,x \ge {\xi _n} \\
 \end{array} \right.,
\end{align}
\emph{where we have ${\zeta _n} = \frac{{{a_n}{P_S}{N_A}}}{{R_{{D_1}}^\alpha  + {a_m}{P_A} + 1}}$ and ${\xi _n} = \frac{{{a_n}{P_S}{N_A}}}{{{a_m}{P_A} + 1}}$.}
\begin{proof}
 Based on  \eqref{SINR n}, we can express the asymptotic CDF of   $F_{{B_n},\infty }^{AN}$ as $F_{{B_n},\infty }^{AN}\left( x \right) = \Pr \left\{ {\frac{{{a_n}{P_S}{N_A}}}{{{a_m}{P_A} + 1 + d_n^\alpha }} \le x} \right\}$.
After some further mathematical manipulations, we can obtain the CDF of $F_{{B_n},\infty }^{AN}$ for large antenna arrays. The proof is completed.
\end{proof}
\end{lemma}
We then derive the asymptotic CDF of user $m$ for ${N_A} \to \infty$.
\begin{lemma}\label{lemma:CDF Bm large final}
\emph{Assuming that user $m$ is randomly located in the ring $D_2$ of Fig.~\ref{system_model_AN} and ${N_A} \to \infty$, the CDF of $F_{{B_m},\infty }^{AN}$ is given by}
\begin{align}\label{CDF Bm large final}
F_{{B_m},\infty }^{AN}\left( x \right) = \left\{ \begin{array}{l}
 1,x \ge {\zeta _{m1}} \\
 \frac{{R_{{D_2}}^2 - t_m^2 + {b_1}{e^{ - \frac{{{a_m}{P_S}{N_A}}}{{x{a_n}{P_S}}}}}}}{{R_{{D_2}}^2 - R_{{D_1}}^2}} \\
  \times \int_{{R_{{D_1}}}}^{{t_m}} r {e^{\frac{{{r^\alpha }}}{{{a_n}{P_S}}}}}dr,{\zeta _{m2}} < x \le {\zeta _{m1}} \\
 \frac{{{b_1}{e^{ - \frac{{{a_m}{P_S}{N_A}}}{{x{a_n}{P_S}}}}}}}{{R_{{D_2}}^2 - R_{{D_1}}^2}}\int_{{R_{{D_1}}}}^{{R_{{D_2}}}} r {e^{\frac{{{r^\alpha }}}{{{a_n}{P_S}}}}}dr,x < {\zeta _{m2}} \\
 \end{array} \right.,
\end{align}
\emph{where we have $b_1 = 2{e^{\frac{{{a_n}{P_A} + 1}}{{{a_n}{P_S}}}}}$, ${t_m} = \sqrt[\alpha ]{{\frac{{{a_m}{P_S}{N_A}}}{x} - {a_n}{P_A} - 1}}$, ${\zeta _{m1}} = \frac{{{a_m}{P_S}{N_A}}}{{R_{{D_1}}^\alpha  + {a_n}{P_A} + 1}},{\zeta _{m2}} = \frac{{{a_m}{P_S}{N_A}}}{{R_{{D_2}}^\alpha  + {a_n}{P_A} + 1}}$, and ${\xi _m} = \frac{{{a_m}{P_S}{N_A}}}{{{a_n}{P_A} + 1}}.$}
\begin{proof}
Similarly, based on  \eqref{SINR m},the CDF of the asymptotic $F_{{B_m},\infty }^{AN}$ is given by
\begin{align}\label{CDF Bm large 1}
F_{{B_m},\infty }^{AN}\left( x \right) = \Pr \left\{ {\frac{{{a_m}{P_S}{N_A}}}{{{a_n}{P_S}{{\left| {{{\bf{h}}_m}\frac{{{\bf{h}}_n^\dag }}{{\left\| {{{\bf{h}}_n}} \right\|}}} \right|}^2} + {a_n}{P_A} + 1 + d_m^\alpha }} \le x} \right\}.
\end{align}
After some further mathematical manipulations, we obtain the CDF of $F_{{B_m},\infty }^{AN}$ for large antenna arrays. The proof is completed.
\end{proof}
\end{lemma}
Let us now turn our attention to the derivation of the Eves' PDF in a large-scale antenna scenario.
\begin{lemma}\label{lemma:PDF Ek large}
\emph{Assuming that the Eves distribution obeys a PPP and that AN is generated at the BS, the Eve-exclusion zone has a radius of $r_p$, and ${N_A} \to \infty$, the PDF of ${f_{\gamma _{{E_\kappa },\infty }^{AN}}}$ (where ${\kappa  \in \left\{ {m,n} \right\}}$) is given by}
\begin{align}\label{PDF Ek large}
&{f_{\gamma _{{E_\kappa },\infty }^{AN}}}\left( x \right) = {e^{ - \frac{{\mu _{\kappa 1}^{AN}\Gamma \left( {\delta ,\mu _{\kappa 2}^{AN}x} \right){e^{ - \frac{{{P_A}x}}{{{a_\kappa }{P_S}}}}}}}{{{x^\delta }}} - \frac{{{P_A}x}}{{{a_\kappa }{P_S}}}}}\mu _{\kappa 1}^{AN}{x^{ - \delta }}\nonumber\\
&\times\left( {{{\left( {\mu _{\kappa 2}^{AN}} \right)}^\delta }{x^{\delta  - 1}}{e^{ - \mu _{\kappa 2}^{AN}x}} + \Gamma \left( {\delta ,\mu _{\kappa 2}^{AN}x} \right)\left( {\frac{{{P_A}}}{{{a_\kappa }{P_S}}} + \frac{\delta }{x}} \right)} \right).
\end{align}
\begin{proof}
Using the theorem of large values, we have $\mathop {\lim }\limits_{{N_A} \to \infty } I_{e,\infty }^{AN} = {a_m}\sigma _a^2{\left\| {{{\bf{h}}_e}{{\bf{V}}_m}} \right\|^2} + {a_n}\sigma _a^2{\left\| {{{\bf{h}}_e}{{\bf{V}}_n}} \right\|^2} \to {P_A}$. The asymptotic CDF of ${F_{\gamma _{{E_\kappa },\infty }^{AN}}}$ associated with ${N_A} \to \infty $ is given by
\begin{align}\label{CDF Ek large 1}
&{F_{\gamma _{{E_\kappa },\infty }^{AN}}}\left( x \right) = \Pr \left\{ {\mathop {\max }\limits_{e \in {\Phi _e},{d_e} \ge {r_p}} \left\{ {\frac{{{a_\kappa }{P_S}{X_{e,\kappa }}}}{{I_{e,\infty }^{AN} + d_e^\alpha }}} \right\} \le x} \right\}\nonumber\\
&= {E_{{\Phi _e}}}\left\{ {\prod\limits_{e \in {\Phi _e},{d_e} \ge {r_p}} {{F_{{X_{e,\kappa }}}}\left( {\frac{{\left( {{P_A} + d_e^\alpha } \right)x}}{{{a_\kappa }{P_S}}}} \right)} } \right\}.
\end{align}
Following the procedure used for deriving \eqref{gamma XE_CDF 2}, we apply the generating function and switch to polar coordinates. Then with the help of \cite[ Eq. (3.381.9)]{gradshteyn}, \eqref{CDF Ek large 1} can be expressed as
\begin{align}\label{CDF Ek large 2}
{F_{\gamma _{{E_\kappa },\infty }^{AN}}}\left( x \right) = \exp \left[ { - \frac{{\mu _{\kappa 1}^{AN}\Gamma \left( {\delta ,\mu _{\kappa 2}^{AN}x} \right)}}{{{x^\delta }}}{e^{ - \frac{{{P_A}x}}{{{a_\kappa }{P_S}}}}}} \right].
\end{align}
Taking derivative of \eqref{CDF Ek large 2}, we obtain the PDF of ${f_{\gamma _{{E_\kappa },\infty }^{AN}}}$. The proof is completed.
\end{proof}
\end{lemma}
\begin{remark}\label{remark:PDF Ek large}
The results derived in \eqref{PDF Ek large} show that the PDF of ${f_{\gamma _{{E_\kappa },\infty }^{AN}}}$ is independent of the number of antennas $N_A$ in our large antenna array analysis. 
\end{remark}


Let us now derive the SOP for our large antenna array scenario in the following two Theorems.

\begin{corollary}\label{theorem:SOP AN large}
\emph{Assuming that the LUs and Eves distribution obey PPPs, AN is generated at the BS and ${N_A} \to \infty$, the SOP for user $n$ is given by}
\begin{align}\label{CDF Bn AN large}
&P_{n,\infty }^{AN}\left( {{R_n}} \right) = 1 - {e^{ - \frac{{\mu _{n1}^{AN}\Gamma \left( {\delta ,\mu _{n2}^{AN}{\chi _{n2}}} \right)}}{{{{\left( {{\chi _{n2}}} \right)}^\delta }}}{e^{ - \frac{{{P_A}{\chi _{n2}}}}{{{a_n}{P_S}}}}}}}  \nonumber\\
&+ \mu _{n1}^{AN}\int_{{\chi _{n1}}}^{{\chi _{n2}}} {{e^{ - \frac{{\mu _{n1}^{AN}\Gamma \left( {\delta ,\mu _{n2}^{AN}x} \right){e^{ - \frac{{{P_A}x}}{{{a_n}{P_S}}}}}}}{{{x^\delta }}} - \frac{{{P_A}x}}{{{a_n}{P_S}}}}}} {\Xi _2}\nonumber\\
&\times \left( {1 - \frac{1}{{R_{{D_1}}^2}}{{\left( {\frac{{{a_n}{P_S}{N_A}}}{{{2^{{R_n}}}\left( {1 + x} \right) - 1}} - {a_m}{P_A} - 1} \right)}^\delta }} \right)dx,
\end{align}
\emph{where ${\chi _{n1}} = \frac{{{\zeta _n} + 1}}{{{2^{{R_n}}}}} - 1$, ${\chi _{n2}} = \frac{{{\xi _n} + 1}}{{{2^{{R_n}}}}} - 1$, and ${\Xi _2} = {x^{ - \delta }}\left( {{{\left( {\mu _{n2}^{AN}} \right)}^\delta }{{{x^{\delta  - 1}}}}{e^{ - \mu _{n2}^{AN}x}} + \Gamma \left( {\delta ,\mu _{n2}^{AN}x} \right)\left( {\frac{{{P_A}}}{{{a_n}{P_S}}} + \frac{\delta }{x}} \right)} \right)$.}
\begin{proof}
Using the results of \textbf{Lemma \ref{lemma:CDF Bn large final}} and \textbf{Lemma \ref{lemma:PDF Ek large}}, upon substituting \eqref{CDF Bn large 2} and \eqref{PDF Ek large} into \eqref{SOP n}, we can express the SOP for user $n$.
\end{proof}
\end{corollary}

\begin{corollary}\label{theorem:SOP AN large}
\emph{Assuming that the LUs and Eves distribution obey PPPs, AN is generated at the BS,  and ${N_A} \to \infty$, the SOP for user $m$ is given by \eqref{CDF Bm AN large} at the top of the next page,}
\begin{figure*}[!t]
\normalsize
\begin{align}\label{CDF Bm AN large}
&P_{m,\infty }^{AN}\left( {{R_m}} \right) = 1 - {e^{ - \frac{{\mu _{\kappa 1}^{AN}\Gamma \left( {\delta ,\mu _{\kappa 2}^{AN}{\chi _{m1}}} \right)}}{{{{\left( {{\chi _{m1}}} \right)}^\delta }}}{e^{ - \frac{{{P_A}{\chi _{m1}}}}{{{a_\kappa }{P_S}}}}}}} + \frac{{\mu _{m1}^{AN}{b_1}{\Lambda _1}}}{{R_{{D_2}}^2 - R_{{D_1}}^2}}\int_0^{{\chi _{m2}}} {{e^{ - \frac{{\mu _{m1}^{AN}\Gamma \left( {\delta ,\mu _{m2}^{AN}x} \right){e^{ - \frac{{{P_A}x}}{{{a_m}{P_S}}}}}}}{{{x^\delta }}} - \frac{{{a_m}{P_S}{N_A}}}{{\left( {{2^{{R_m}}}\left( {1 + x} \right) - 1} \right){a_n}{P_S}}} - \frac{{{P_A}x}}{{{a_m}{P_S}}}}}{\Xi _1}dx}  \nonumber\\
&+ \frac{{\mu _{m1}^{AN}}}{{R_{{D_2}}^2 - R_{{D_1}}^2}}\int_{{\chi _{m2}}}^{{\chi _{m1}}} {{e^{ - \frac{{\mu _{m1}^{AN}\Gamma \left( {\delta ,\mu _{m2}^{AN}x} \right){e^{ - \frac{{{P_A}x}}{{{a_m}{P_S}}}}}}}{{{x^\delta }}} - \frac{{{P_A}x}}{{{a_m}{P_S}}}}}} \left( {R_{{D_2}}^2 - t_{m*}^2 + {b_1}{e^{ - \frac{{{a_m}{P_S}{N_A}}}{{\left( {{2^{{R_m}}}\left( {1 + x} \right) - 1} \right){a_n}{P_S}}}}}} \right){\Xi _1}{\Lambda _2}dx,
\end{align}
\hrulefill \vspace*{0pt}
\end{figure*}
\emph{where we have ${\Xi _1} = {x^{ - \delta }}\left( {\mu _{m2}^{AN}{{\left( {\mu _{m2}^{AN}x} \right)}^{\delta  - 1}}{e^{ - \mu _{m2}^{AN}x}} + \Gamma \left( {\delta ,\mu _{m2}^{AN}x} \right)\left( {\frac{{{P_A}}}{{{a_m}{P_S}}} + \frac{\delta }{x}} \right)} \right)$, ${\Lambda _1} = \int_{{R_{{D_1}}}}^{{R_{{D_2}}}} {r{e^{\frac{{{r^\alpha }}}{{{a_n}{P_S}}}}}dr} , {\Lambda _2} = \int_{{R_{{D_1}}}}^{{t_{m*}}} {r{e^{\frac{{{r^\alpha }}}{{{a_n}{P_S}}}}}} dr$, ${t_{m*}} = \sqrt[\alpha ]{{\frac{{{a_m}{P_S}{N_A}}}{{{2^{{R_m}}}\left( {1 + x} \right) - 1}} - {a_n}{P_A} - 1}}$, and ${\chi _{m2}} = \frac{{{\zeta _{m2}} + 1}}{{{2^{{R_m}}}}} - 1$.}
\begin{proof}
 Using the results of \textbf{Lemma~\ref{lemma:CDF Bm large final}} and \textbf{Lemma~\ref{lemma:PDF Ek large}}, upon substituting \eqref{CDF Bm large final} and \eqref{PDF Ek large} into \eqref{SOP m}, we can express the SOP for user $m$. The proof is completed.
\end{proof}
\end{corollary}

\begin{proposition}\label{proposition:SOP AN total}
Under the assumption of ${N_A} \to \infty$, the SOP of multiple-antenna aided scenario relaying on AN for the selected user pair can be expressed as
\begin{align}\label{Pout_total_AN_large}
P^{AN}_{mn,\infty} = 1 - \left( {1 - P^{AN}_{m,\infty}} \right)\left( {1 - P^{AN}_{n,\infty}} \right).
\end{align}
where $P^{AN}_{n,\infty}$ and $P^{AN}_{m,\infty}$ are given by \eqref{CDF Bn AN large} and \eqref{CDF Bm AN large}, respectively.
\end{proposition}

\section{Numerical Results}
In this section, our numerical results are presented for characterizing the performance of large-scale networks.  The complexity-vs-accuracy tradeoff parameter is $K=20$. Table~\ref{parameter} summarizes the the Monte Carlo simulation parameters used in this section. BPCU is short for bit per channel use.


\begin{table}[!t]
\centering
\caption{Table of Parameters}
\label{parameter}
\begin{tabular}{|l|l|}
\hline
Monte Carlo simulations repeated  &  ${10^6}$ times \\ \hline
The radius of a disc region for Eves  &  $1000$ m \\ \hline
power sharing coefficients of NOMA&  $a_m=0.6$, $a_n=0.4$   \\ \hline
Targeted secrecy rates & $R_m=R_n=0.1$ BPCU  \\ \hline
Pass loss exponent  & $\alpha=4$  \\ \hline
The radius of the user zone of Section II &  ${R_{{D}}}=10$ m \\ \hline
The radius of the user zone of Section III & $R_{D_1}=5$ m, $R_{D_2}=10$ m \\ \hline
\end{tabular}
\end{table}

\subsection{Secrecy outage probability with channel ordering}
\begin{figure} [t!]
\centering
\includegraphics[width= 3.8in, height=2.6in]{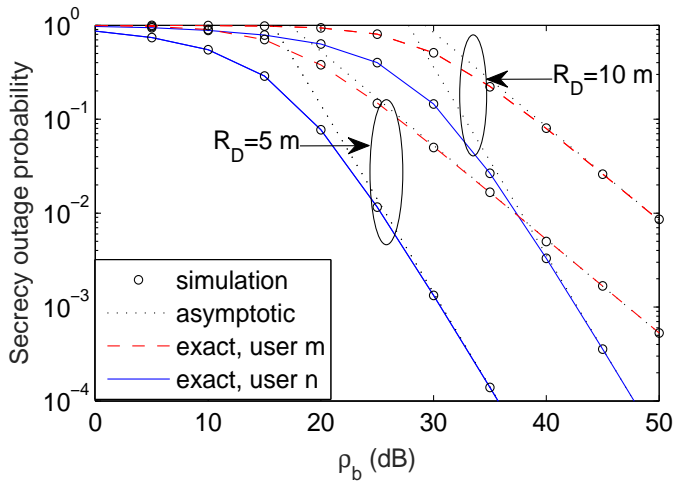}
 \caption{The SOP versus $\rho_b$, with $\rho_e=10$~dB, $\alpha=4$, $\lambda_e= 10^{-3}$, $M=2$, $m=1$, $n=2$, and $r_p=10$~m. The exact analytical results are calculated from \eqref{Pout_m} and \eqref{Pout_n}. The asymptotic analytical results are calculated from \eqref{SOP n asym} and \eqref{SOP m asym}.}\label{SOP_asym}
\end{figure}

\begin{figure} [t!]
\centering
\includegraphics[width= 3.8in, height=2.6in]{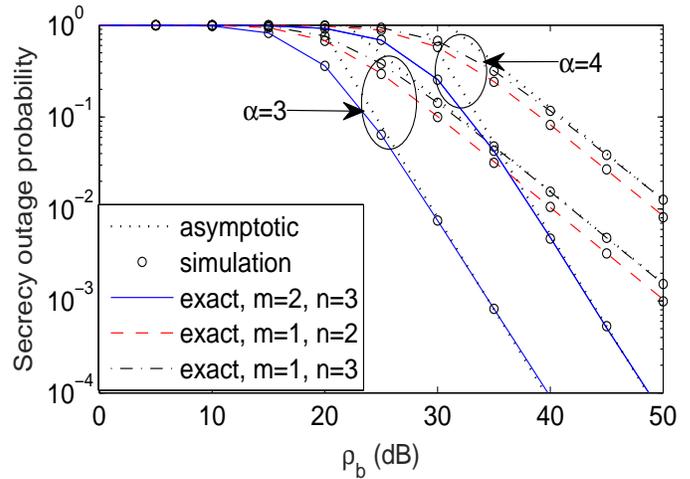}
 \caption{The SOP of  user pair versus $\rho_b$, with $\rho_e=10$ dB, $\lambda_e= 10^{-3}$, $R_D=10$ m, $M=3$, and $r_p=10$ m. The exact analytical results are calculated from \eqref{Pout_total}. The asymptotic analytical results are calculated from \eqref{Pout_total asy}.}\label{SOP_pair}
\end{figure}

\begin{figure} [t!]
\centering
\includegraphics[width= 3.8in, height=2.6in]{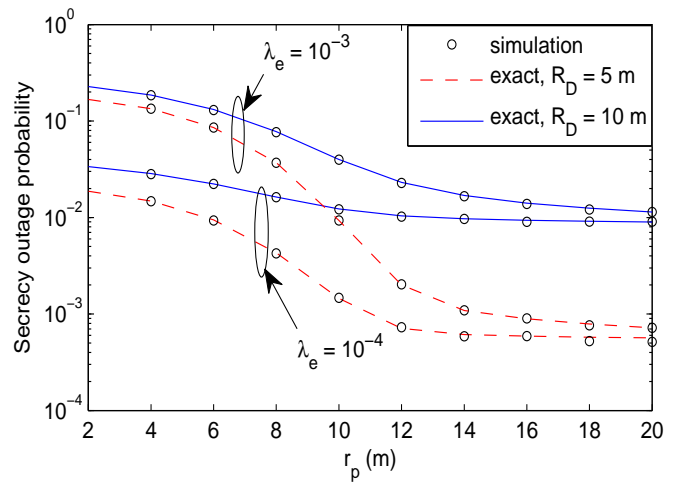}
 \caption{The SOP of  user pair versus $r_p$, with $\rho_b=50$ dB, $\rho_e=40$ dB, $M=2$, $m=1$, $n=2$, and $\alpha=4$. The exact analytical results are calculated from \eqref{Pout_total}.}\label{SOP_pair_zone}
\end{figure}

From Fig. \ref{SOP_asym} to Fig. \ref{SOP_pair_zone}, we investigate the secrecy performance in conjunction with channel ordering, which correspond to the scenario considered in Section II.

Fig. \ref{SOP_asym} plots the SOP of a single user ($m$-th and $n$-th) versus $\rho_b$ for different user zone radii. The curves represent the exact analytical SOP of both the $m$-th user and of $n$-th user derived in \eqref{Pout_m} and \eqref{Pout_n}, respectively. The asymptotic analytical SOP  of both the $m$-th and $n$-th users, are derived in \eqref{SOP m asym} and \eqref{SOP n asym}, respectively.  Fig.~\ref{SOP_asym} confirms the close agreement between the simulation and analytical results. A specific observation is that the reduced SOP can be achieved by reducing the radius of the user zone, since a smaller user zone leads to a lower path-loss. Another observation is that the $n$-th user has a more steep slope than the $m$-th user. This is due to the fact that we have $m<n$ and the $m$-th user as well as $n$-th user achieve a secrecy diversity order of $m$ and $n$ respectively, as inferred from \eqref{SOP m asym} and \eqref{SOP n asym}.

Fig. \ref{SOP_pair} plots the SOP of the selected user pair versus the transmit SNR $\rho_b$ for different path-loss factors. The  exact analytical SOP curves are plotted from \eqref{Pout_total}. The  asymptotic analytical SOP curves are plotted from \eqref{Pout_total asy}. It can be observed that the two kinds of dashed curves have the same slopes. By contrast, the solid curves indicate a higher secrecy outage slope, which is due to the fact that the secrecy diversity order of the user pair is determined by that of the poor one. This phenomenon is also confirmed by the insights in \textbf{Remark 1}.

Fig. \ref{SOP_pair_zone} plots the SOP of the selected user pair versus $r_p$ for different densities of the Eves. We can observe that as expected, the SOP decreases, as the radius of the Eve-exclusion zone increases. Another option for enhancing the PLS is to reduce the radius of the user zone, since it reduces the total path loss. It is also worth noting that having a lower E density $\lambda_e$ results in an improved PLS, i.e. reduced SOP. This behavior is due to the plausible fact that a lower $\lambda_e$ results in having less Eves, which degrades the multiuser diversity gain, when the most detrimental E is selected. As a result, the destructive capability of the most detrimental E is reduced and hence the SOP is improved. It is worth pointing out that dynamic power sharing between two users is capable of improving the secrecy performance of  the scenarios considered, but this is  beyond the scope of this paper.
\subsection{Secrecy outage probability with artificial noise}
\begin{figure} [t!]
\centering
\includegraphics[width= 3.8in, height=2.6in]{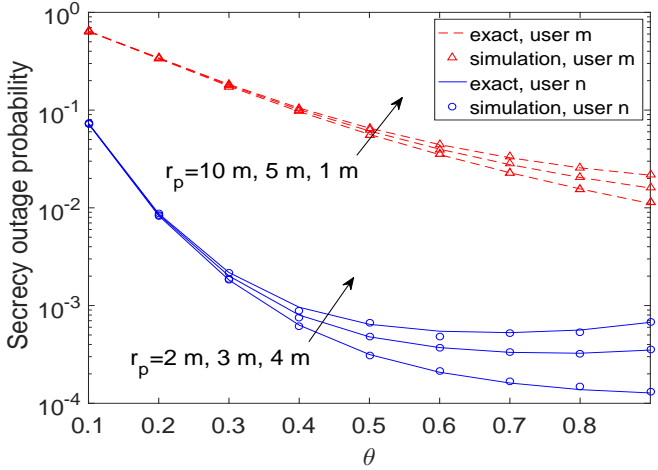}
 \caption{The SOP versus $\theta$, with $\alpha=4$, $R_{D_1}=5$~m, $R_{D_2}=10$~m, $\lambda_e= 10^{-4}$, $N_A=4$, $\rho_t=30$ dB.  The exact analytical results are calculated from ~\eqref{SOP m AN} and~\eqref{SOP n AN}.}\label{ana_rp=5_lamda10-4_legend}
\end{figure}

\begin{figure} [t!]
\centering
\includegraphics[width= 3.8in, height=2.6in]{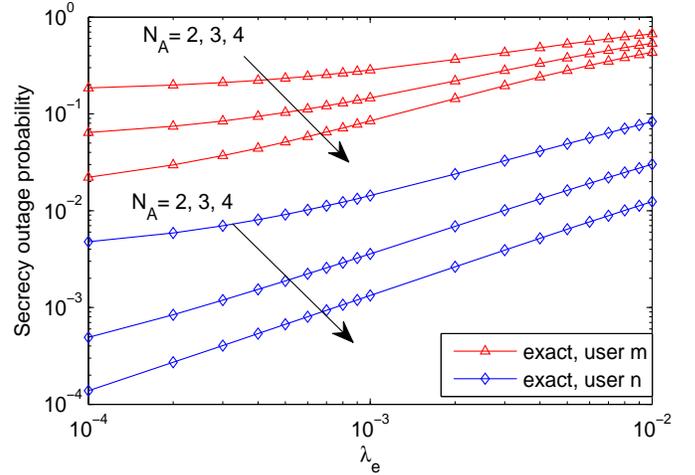}
 \caption{The SOP versus $\lambda_e$, with $\theta=0.8$, $\alpha=4$, $R_{D_1}=5$~m, $R_{D_2}=10$~m, $\rho_t=30$~dB, $r_p=4$~m. The exact analytical results are calculated from ~\eqref{SOP m AN} and~\eqref{SOP n AN}.}\label{P_out_antenna_different_lambda}
\end{figure}

\begin{figure} [t!]
\centering
\includegraphics[width= 3.8in, height=2.6in]{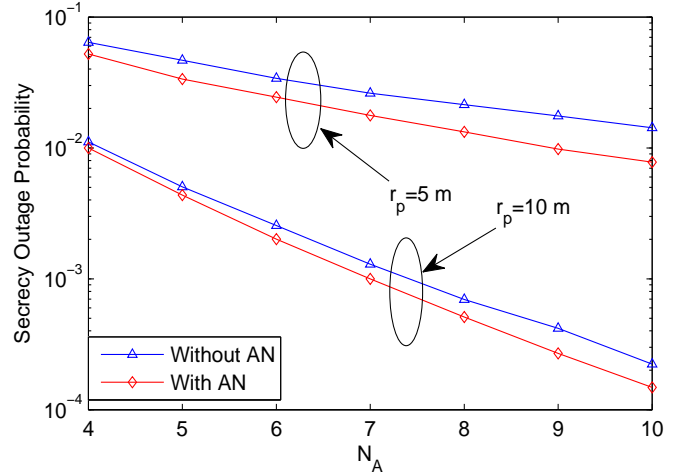}
 \caption{The SOP of the user pair versus $N_A$, with $R_{D_1}=5$~m, $R_{D_2}=10$~m, $\alpha=3$, $\lambda_e= 10^{-3}$, $\rho_t=30$~dB. }\label{P_out_alpha_different_antenna}
\end{figure}

\begin{figure} [t!]
\centering
\includegraphics[width= 3.8in, height=2.6in]{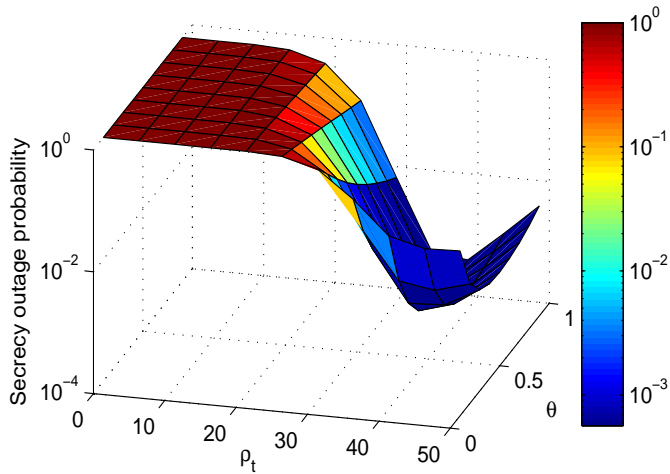}
 \caption{SOP of the user pair versus $\rho_t$ and $\theta$, with $N_A=4$, $\alpha=4$, $R_{D_1}=5$~m, $R_{D_2}=10$~m, $\lambda_e= 10^{-4}$, $r_p=10$~m. The exact analytical results are calculated from ~\eqref{Pout_total AN}.}\label{P_out_mn_3D_final}
\end{figure}

\begin{figure} [t!]
\centering
\includegraphics[width= 3.8in, height=2.6in]{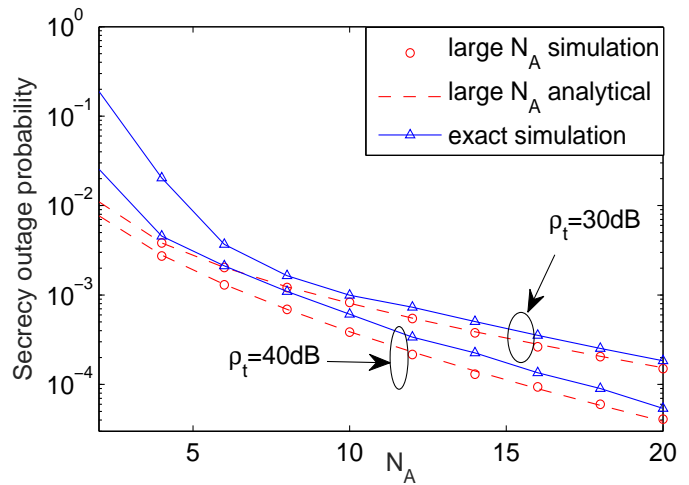}
 \caption{Large analysis for the SOP of user pair versus $N_A$, with $\theta=0.8$, $R_{D_1}=5$~m, $R_{D_2}=10$~m, $\lambda_e= 10^{-4}$, $r_p=5$~m. The asymptotic analytical results are calculated from ~\eqref{Pout_total_AN_large}.}\label{P_out_large}
\end{figure}

From Fig. \ref{ana_rp=5_lamda10-4_legend} to Fig. \ref{P_out_large}, we  investigate the secrecy performance in the presence of AN, which correspond to the scenario considered in Section III.

Fig.~\ref{ana_rp=5_lamda10-4_legend} plots the SOP of user $m$ and user $n$ versus $\theta$ for different Eve-exclusion zones. The solid and dashed curves represent the analytical performance of user $m$ and user $n$, corresponding to the results derived in~\eqref{SOP m AN} and~\eqref{SOP n AN}. Monte Carlo simulations are used for verifying our derivations. Fig.~\ref{ana_rp=5_lamda10-4_legend} confirms a close agreement between the simulation and analytical results. Again, a reduced SOP can be achieved by increasing the Eve-exclusion zone, which degrades the channel conditions of the Eves. Another observation is that user $n$ achieves a lower SOP than user $m$, which is explained as follows: 1) user $n$ has better channel conditions than user $m$, owing to its lower path loss; and 2) user $n$ is capable of cancelling the interference imposed by user $m$ using SIC techniques, while user $m$ suffers from the interference inflicted by user $n$. It is also worth noting that the SOP is not a monotonic function of $\theta$. This phenomenon indicates that there exists an optimal value for power allocation, which depends on the system parameters.

Fig. \ref{P_out_antenna_different_lambda} plots the SOP of user $m$ and user $n$ versus $\lambda_e$ for different number of antennas. We can observe that the SOP decreases, as the E density is reduced. This behavior is caused by the fact that a lower $\lambda_e$ leads to having less Eves, which reduces the multiuser diversity gain, when the most detrimental E is considered. As a result, the distinctive capability of the most detrimental E is reduced and hence the secrecy performance is improved. It is also worth noting that increasing the number of antennas is capable of increasing the secrecy performance. This is due to the fact that ${{\left\| {{{\bf{h}}_m}} \right\|}^2}$ in \eqref{SINR m} and ${{\left\| {{{\bf{h}}_n}} \right\|}^2}$ in \eqref{SINR n}  both follow $Gamma\left( {{N_A},1} \right)$ distributions, which is the benefit of the improved multi-antenna diversity gain.

Fig. \ref{P_out_alpha_different_antenna} plots the SOP of the selected user pair versus $N_A$ for different path loss exponents. In this figure, the curves representing the case without AN are generated by setting $\theta=1$, which means that all the power is allocated to the desired signal. In this case, the BS only uses beamforming for transmitting the desired signals and no AN is generated. The curves in the presence of AN are generated  by setting $\theta=0.9$. We show that the PLS can be enhanced by using AN. This behavior is caused by the fact that at the receiver side, user $m$ and user $n$ are only affected by the AN generated by each other; By contrast, the Eves are affected by the AN of both user $m$ and user $n$. We can observe that the SOP of the selected user pair decreases, as the  Eve-exclusion radius increases.

Fig. \ref{P_out_mn_3D_final} plots the SOP of the selected user pair versus $\rho_t$ and $\theta$. It is observed that the SOP first decreases then increases as $\rho_t$ increases, which is in contrast to the traditional trend, where the SOP always decreases as the transmit SNR increases. This behavior can be explained as follows. The SOP of the selected user pair is determined by user $m$. As $\rho_t$ increases, on the one hand, the signal power of user $m$ is increased, which improves the secrecy performance; On the other hand, user $m$ also suffers from the interference imposed by user $n$ (including both the signal and AN), because when $\rho_t$ increases,  the signal power of user $n$ is also increased, which in turn degrades the secrecy performance. As a consequence, there is a tradeoff between $\rho_t$ and the SOP.  It is also noted that the power sharing factor $\theta$ also affect the optimal SOP associated with different values of  $\rho_t$. This phenomenon indicates that it is of salient significance to select beneficial system parameters. Furthermore, optimizing the parameters $\rho_t$ and $\theta$ is capable of further improving the SOP.

Fig. \ref{P_out_large} plots the SOP of large antenna arrays of the selected user pair versus $N_A$ parameterized by different transmit SNRs. The dashed curves represent the analytical SOP of the selected user pair, corresponding to the results derived in~\eqref{Pout_total_AN_large}. We observe a close  agreement between the theoretical analysis and the Monte Carlo simulations, which verifies the accuracy of our derivations. We observe that as $N_A$ increases, the  approximation used in our analysis approaches  the exact SOP. This phenomenon indicates that the  asymptotic SOP derived converges to the exact values, when ${N_A}$ is a sufficiently large number.

\section{Conclusions}
In this paper, the secrecy performance of applying the NOMA protocol in large-scale networks was examined. Specifically, stochastic geometry based techniques were used for modeling both the locations of NOMA users and of the Eves in the  networks considered. Additionally, new analytical SOP expressions were derived for characterizing the system's secrecy performance in both single-antenna and multiple-antenna scenarios. For the single-antenna scenario, the secrecy diversity order of the user pair was also characterized. It was analytically demonstrated that the secrecy diversity order was determined by that one of the user pair who had a poorer channel. For the multiple-antenna scenario, it was shown that the Eves' channel quality is independent of the number of antennas at the BS for large antenna array scenarios. Numerical results were also presented for validating the analysis. It was concluded that the secrecy performance can be improved both by extending the Eve-exclusion zone and by generating AN at the BS. Assuming perfect SIC operations may lead to overestimating the performance of  the networks considered, hence our future research may consider investigating imperfect SIC.  Optimizing the power sharing between two NOMA users is capable of further improving the secrecy performance of the networks considered, which is another promising future research direction.

\numberwithin{equation}{section}
\section*{Appendix~A: Proof of Lemma~\ref{lemma:gamma Bn_CDF}} \label{Appendix:A}
\renewcommand{\theequation}{A.\arabic{equation}}
\setcounter{equation}{0}

To derive the CDF of ${F_{{\gamma _B}}}$, based on \eqref{SNR B n}, we can formulate
\begin{align}\label{gamma Bn_CDF1}
{F_{{\gamma _B}}}\left( x \right) = \Pr \left\{ {\rho_b {a_n}{{\left| {{h_n}} \right|}^2} \le x} \right\} = {F_{{{\left| {{h_n}} \right|}^2}}}\left( {\frac{x}{{\rho_b {a_n}}}} \right),
\end{align}
where ${F_{{{\left| {{h_n}} \right|}^2}}}$ is the CDF of the ordered channel gain for the $n$-th user.
Assuming $y = \frac{x}{{\rho_b {a_n}}}$, and using order statistics~\cite{order}  as well as applying binary series expansion, the CDF of the ordered channels has a relationship with the unordered channels captured as follows:
\begin{align}\label{gamma Bn_CDF2}
{F_{{{\left| {{h_n}} \right|}^2}}}\left( y \right) = {\varphi _n}\sum\limits_{p = 0}^{M - n} {M - n \choose
 p} \frac{{{{\left( { - 1} \right)}^p}}}{{n + p}}{\left( {{F_{{{\left| {{{\tilde h}_n}} \right|}^2}}}\left( y \right)} \right)^{n + p}},
\end{align}
where ${{F_{{{\left| {{{\tilde h}_n}} \right|}^2}}}}$ is the CDF of unordered channel gain for the $n$-th user.

Based on the assumption of homogeneous PPP, and by relying on polar coordinates, ${{F_{{{\left| {{{\tilde h}_n}} \right|}^2}}}}$ is expressed as
\begin{align}\label{gamma Bn_unordered 1}
{F_{{{\left| {{{\tilde h}_n}} \right|}^2}}}\left( y \right) = \frac{2}{{R_D^2}}\int_0^{{R_D}} {\left( {1 - {e^{ - \left( {1 + {r^\alpha }} \right)y}}} \right)rdr}.
\end{align}
However, it is challenging to arrive at an easily implemented insightful expression for ${F_{{{\left| {{{\tilde h}_n}} \right|}^2}}}\left( y \right)$. Therefore, the Gaussian-Chebyshev quadrature relationship~\cite{yuanwei2016TVT} is invoked for finding an approximation of \eqref{gamma Bn_unordered 1} in the following form:
\begin{align}\label{gamma Bn_unordered GC}
{F_{{{\left| {{{\tilde h}_n}} \right|}^2}}}\left( y \right) \approx \sum\limits_{k = 0}^K {{b_k}{e^{ - {c_k}y}}}.
\end{align}

Substituting \eqref{gamma Bn_unordered GC} into \eqref{gamma Bn_CDF2} and applying the multinomial theorem, the CDF ${F_{{{\left| {{h_n}} \right|}^2}}}$  of ordered channel gain is given by
\begin{align}\label{gamma Bn_ordered GC}
&{F_{{{\left| {{h_n}} \right|}^2}}}\left( y \right) \approx  {\varphi _n}\sum\limits_{p = 0}^{M - n} {
 M - n \choose
 p } \frac{{{{\left( { - 1} \right)}^p}}}{{n + p}}\nonumber\\
  &\times{\sum _{\tilde S_n^p}} {{
 n + p \choose
 {q_0} +  \cdots  + {q_K} }\left( {\prod\limits_{k = 0}^K {b_k^{{q_k}}} } \right){e^{ - \sum\limits_{k = 0}^K {{q_k}{c_k}y} }}} .
\end{align}

Substituting $y = \frac{x}{{\rho_b {a_n}}}$ into \eqref{gamma Bn_ordered GC}, we can obtain \eqref{gamma Bn_CDF}. The proof is completed.

\section*{Appendix~B: Proof of Lemma~\ref{lemma:CDF Bn AN}} \label{Appendix:B}
\renewcommand{\theequation}{B.\arabic{equation}}
\setcounter{equation}{0}
Based on \eqref{SINR n}, we express the CDF of $F_{{B_n}}^{AN}$ as follows:
\begin{align}\label{CDF_Bn_AN_1}
&F_{{B_n}}^{AN}\left( x \right) = \Pr \left\{ {{{\left\| {{{\bf{h}}_n}} \right\|}^2} \le x\vartheta \left( {\frac{{{P_A}}}{{{N_A} - 1}}{Y_n} + \frac{{1 + d_n^\alpha }}{{{a_m}}}} \right)} \right\}\nonumber\\
&= 1 - \sum\limits_{p = 0}^{{N_A} - 1} {\frac{{{\vartheta ^p}{x^p}}}{{p!}}} \sum\limits_{q = 0}^p {
 p \choose
 q } {Q_3}\nonumber\\
 &\times \int\limits_{{D_1}} {{e^{ - \frac{{\vartheta x}}{{{a_m}}}\left( {1 + d_n^\alpha } \right)}}{{\left( {\frac{1}{{{a_m}}}\left( {1 + d_n^\alpha } \right)} \right)}^{p - q}}{f_{{D_1}}}\left( {{\omega _n}} \right)d{\omega _n}} ,
\end{align}
where $\vartheta  = \frac{{{a_m}}}{{{a_n}{P_S}}}$, ${Q_3} = \int_0^\infty  {{e^{ - \vartheta x{z_n}}}z_n^q} {f_{I_n^{AN}}}\left( {{z_n}} \right)d{z_n}$, ${f_{I_n^{AN}}}$ and ${f_{{D_1}}}\left( {{\omega _n}} \right)$ are the PDF of $I_n^{AN}$ and $D_1$. Here $I_n^{AN} = \frac{{{P_A}}}{{{N_A} - 1}}{Y_n},{Y_n} = {\left\| {{{\bf{h}}_n}{{\bf{V}}_m}} \right\|^2}$, and ${f_{{D_1}}}\left( {{\omega _n}} \right) = \frac{1}{{\pi R_{{D_1}}^2}}$. Upon changing to polar coordinates and applying \cite[ Eq. (3.381.8)]{gradshteyn}, we arrive at
\begin{align}\label{CDF_Bn_AN_2}
F_{{B_n}}^{AN}\left( x \right) =& 1 - \frac{{\delta {e^{ - \frac{{\vartheta x}}{{{a_m}}}}}}}{{R_{{D_1}}^2}}\sum\limits_{p = 0}^{{N_A} - 1} {\frac{{{\vartheta ^p}{x^p}}}{{p!}}} \sum\limits_{q = 0}^p {
 p \choose
 q } {Q_3}a_m^{q - p}\nonumber\\
 &\times \sum\limits_{u = 0}^{p - q} {
 p - q \choose u } \frac{{\gamma \left( {u + \delta ,\frac{{\vartheta x}}{{{a_m}}}R_{{D_1}}^\alpha } \right)}}{{{{\left( {\frac{{\vartheta x}}{{{a_m}}}} \right)}^{u + \delta }}}}.
\end{align}

Finally we turn our attention on $Q_3$. It is readily seen that $I_n^{AN}$ obeys the Gamma distribution in conjunction with the parameter $\left( {{N_A} - 1,{\frac{{{P_A}}}{{{N_A} - 1}}}} \right)$. Then we can obtain the PDF of ${f_{I_n^{AN}}}\left( {{z_n}} \right) = \frac{{z_n^{{N_A} - 2}{e^{ - \frac{{{z_n}\left( {{N_A} - 1} \right)}}{{{P_A}}}}}}}{{{{\left( {\frac{{{P_A}}}{{{N_A} - 1}}} \right)}^{{N_A} - 1}}\Gamma \left( {{N_A} - 1} \right)}}$. Applying \cite[ Eq. (3.326.2)]{gradshteyn}, we can express $Q_3$ as ${Q_3} = \frac{{\Gamma \left( {{N_A} - 1 + q} \right)}}{{\Gamma \left( {{N_A} - 1} \right){{\left( {\frac{{{P_A}}}{{{N_A} - 1}}} \right)}^{{N_A} - 1}}{{\left( {\vartheta x + \frac{{{N_A} - 1}}{{{P_A}}}} \right)}^{{N_A} - 1 + q}}}}$. Upon substituting $Q_3$ into \eqref{CDF_Bn_AN_2}, we obtain the CDF of $F_{{B_n}}^{AN}\left( x \right)$ as \eqref{CDF Bn AN final}.

\section*{Appendix~C: Proof of Lemma~\ref{lemma:CDF Bm AN}} \label{Appendix:C}
\renewcommand{\theequation}{C.\arabic{equation}}
\setcounter{equation}{0}
Based on~\eqref{SINR m}, we express the CDF of $ F_{{B_m}}^{AN}$ as
\begin{align}\label{CDF Bm AN}
&F_{{B_m}}^{AN}\left( x \right) = \Pr \left\{ {\gamma _{{B_m}}^{AN} \le x} \right\}\nonumber\\
&= \Pr \left\{ {\frac{{{a_m}\sigma _s^2{{\left\| {{{\bf{h}}_m}} \right\|}^2}}}{{{a_n}\sigma _s^2{{\left| {{{\bf{h}}_m}\frac{{{\bf{h}}_n^\dag }}{{\left\| {{{\bf{h}}_n}} \right\|}}} \right|}^2} + {a_n}\sigma _a^2{{\left\| {{{\bf{h}}_m}{{\bf{V}}_n}} \right\|}^2} + 1 + d_m^\alpha }} \le x} \right\}.
\end{align}
It may be readily seen that ${{{\left\| {{{\bf{h}}_m}} \right\|}^2}}$ obeys a Gamma distribution having the parameters of $\left( {{N_A},1} \right)$. Hence the CDF of ${{{\left\| {{{\bf{h}}_m}} \right\|}^2}}$ is given by
\begin{align}\label{CDF hm}
F_{{B_m}}^{AN}\left( x \right) = 1 - {e^{ - x}}\sum\limits_{p = 0}^{{N_A} - 1} {\frac{{{x^p}}}{{p!}}}.
\end{align}
Denoting ${X_m} = {\left| {{{\bf{h}}_m}\frac{{{\bf{h}}_n^\dag }}{{\left\| {{{\bf{h}}_n}} \right\|}}} \right|^2},{Y_m} = {\left\| {{{\bf{h}}_m}{{\bf{V}}_n}} \right\|^2}$, based on \eqref{CDF hm}, we can re-write \eqref{CDF Bm AN} as
\begin{align}\label{CDF Bm AN 1}
&F_{{B_m}}^{AN}\left( x \right) =  \Pr \left\{ {{{\left\| {{{\bf{h}}_m}} \right\|}^2} \le x\nu \left( {I_m^{AN} + \frac{{1 + d_m^\alpha }}{{{a_n}}}} \right)} \right\} \nonumber\\
&= 1 - \int\limits_{{D_2}} {\int_0^\infty  {\sum\limits_{p = 0}^{{N_A} - 1} {\frac{{{{\left( {\nu x\left( {{z_m} + \frac{{1 + d_m^\alpha }}{{{a_n}}}} \right)} \right)}^p}}}{{p!}}} } } \nonumber\\
&  \times \left( {{e^{ - \nu x{z_m} - \nu x\frac{{1 + d_m^\alpha }}{{{a_n}}}}}} \right){f_{I_m^{AN}}}\left( {{z_m}} \right){f_{{D_2}}}\left( {{\omega _m}} \right)d{z_m}d{\omega _m},
\end{align}
where $\nu  = \frac{{{a_n}}}{{{a_m}{P_S}}}$, ${f_{I_m^{AN}}}$ and ${{f_{{D_2}}}}$ are the PDF of $I_m^{AN}$ and ${{D_2}}$, respectively. Here we have $I_m^{AN} = \sigma _s^2{X_m} + \sigma _a^2{Y_m}$ and ${f_{{D_2}}}\left( {{\omega _m}} \right) = \frac{1}{{\pi \left( {R_{{D_2}}^2 - R_{{D_1}}^2} \right)}}$.
Applying a binary series expansion to \eqref{CDF Bm AN 1}, we arrive at:
\begin{align}\label{CDF Bm AN 2}
&F_{{B_m}}^{AN}\left( x \right) = 1 - \sum\limits_{p = 0}^{{N_A} - 1} {\frac{{{\nu ^p}{x^p}}}{{p!}}} \sum\limits_{q = 0}^p {p \choose
 q } {Q_1}\nonumber\\
 &\times \int\limits_{{D_2}} {{e^{ - \nu x\frac{{1 + d_m^\alpha }}{{{a_n}}}}}{{\left( {\frac{{1 + d_m^\alpha }}{{{a_n}}}} \right)}^{p - q}}{f_{{D_2}}}\left( {{\omega _m}} \right)d{\omega _m}}  ,
\end{align}
where ${Q_1} = \int_0^\infty  {{e^{ - \nu x{z_m}}}z_m^q} {f_{I_m^{AN}}}\left( {{z_m}} \right)d{z_m}$. Note that the distance $d_m$ is determined by the location of $\omega _m$. Then we change to polar coordinates and applying a binary series expansion again, we obtain
\begin{align}\label{CDF Bm AN 3}
&F_{{B_m}}^{AN}\left( x \right) = 1 - \frac{{{2e^{ - \frac{{\nu x}}{{{a_n}}}}}}}{{R_{{D_2}}^2 - R_{{D_1}}^2}}\sum\limits_{p = 0}^{{N_A} - 1} {\frac{{{\nu ^p}{x^p}}}{{p!}}} \sum\limits_{q = 0}^p {p \choose q }  \nonumber\\
 &\times {Q_1}\frac{1}{{a_n^{p - q}}}\sum\limits_{u = 0}^{p - q} {
 p - q \choose u } \int_{{R_{{D_1}}}}^{{R_{{D_2}}}} {{r^{u\alpha  + 1}}{e^{ - \nu x{P_S}{r^\alpha }}}} dr.
\end{align}
By invoking \cite[ Eq. (3.381.8)]{gradshteyn}, we obtain
\begin{align}\label{CDF Bm AN 4}
&F_{{B_m}}^{AN}\left( x \right) = 1 - \frac{{{2e^{ - \frac{{\nu x}}{{{a_n}}}}}}}{{R_{{D_2}}^2 - R_{{D_1}}^2}}\sum\limits_{p = 0}^{{N_A} - 1} {\frac{{{\nu ^p}{x^p}}}{{p!}}} \sum\limits_{q = 0}^p {
 p \choose
 q } {Q_1}\frac{1}{{a_n^{p - q}}}\nonumber\\
 &\times\sum\limits_{u = 0}^{p - q} {
 p - q \choose u } \frac{{\gamma \left( {u + \delta ,\frac{{\nu x}}{{{a_n}}}R_{{D_2}}^\alpha } \right) - \gamma \left( {u + \delta ,\frac{{\nu x}}{{{a_n}}}R_{{D_1}}^\alpha } \right)}}{{\alpha {{\left( {\frac{{\nu x}}{{{a_n}}}} \right)}^{u + \delta }}}}.
\end{align}
Let us now turn our attention to the derivation of the integral $Q_1$ in \eqref{CDF Bm AN 2} -- \eqref{CDF Bm AN 4}. Note that $X_m$ follows the exponential distribution with unit mean, while  $Y_m$ follows the distribution ${Y_m} \sim Gamma\left( {{N_A} - 1,1} \right)$. As such, the PDF of ${f_{I_m^{AN}}}$ is given by~\cite{zhang2013enhancing}
\begin{align}\label{PDF Im_AN}
{f_{I_m^{AN}}}\left( {{z_m}} \right) = \left\{ \begin{array}{l}
 \frac{{{t_1}}}{{{e^{\frac{{{z_m}}}{{{P_S}}}}}}}\left( {1 - \sum\limits_{l = 0}^{{N_A} - 2} {\frac{{{{\left( {\frac{{{N_A} - 1}}{{{P_A}}} - \frac{1}{{{P_S}}}} \right)}^l}z_m^l}}{{l!{e^{\left( {\frac{{{N_A} - 1}}{{{P_A}}} - \frac{1}{{{P_S}}}} \right){z_m}}}}}} } \right),\theta  \ne \frac{1}{{{N_A}}} \\
 \frac{{z_m^{{N_A} - 1}{e^{ - \frac{{{z_m}}}{{{P_S}}}}}}}{{{P_S}^{{N_A}}\left( {{N_A} - 1} \right)!}},\theta  = \frac{1}{{{N_A}}} \\
 \end{array} \right.,
\end{align}
where we have ${t_1} = \frac{{{{\left( {1 - \frac{{{P_A}}}{{\left( {{N_A} - 1} \right){P_S}}}} \right)}^{1 - {N_A}}}}}{{{P_S}}}$. Based on \eqref{PDF Im_AN}, and applying \cite[ Eq. (3.326.2)]{gradshteyn}, we can express $Q_1$ as follows:
\begin{align}\label{Q_1_2}
{Q_1} = \left\{ \begin{array}{l}
 {\frac{{t_1}{\Gamma \left( {q + 1} \right)}}{{{{\left( {x\nu  + \frac{1}{{{P_S}}}} \right)}^{q + 1}}}} - \sum\limits_{l = 0}^{{N_A} - 2} {\frac{{\frac{{t_1}}{{l!}}{{\left( {\frac{{{N_A} - 1}}{{{P_A}}} - \frac{1}{{{P_S}}}} \right)}^l}\Gamma \left( {q + l + 1} \right)}}{{{{\left( {\nu x + \frac{{{N_A} - 1}}{{{P_A}}}} \right)}^{q + l + 1}}}}} },\theta  \ne \frac{1}{{{N_A}}} \\
 \frac{{\Gamma \left( {q + {N_A}} \right)}}{{{P_S}^{{N_A} }\left( {{N_A} - 1} \right)!{{\left( {\nu x + \frac{1}{{{P_S}}}} \right)}^{q + {N_A}}}}},\theta  = \frac{1}{{{N_A}}} \\
 \end{array} \right..
\end{align}
Upon substituting \eqref{Q_1_2} into \eqref{CDF Bm AN 4}, the CDF of $F_{{B_m}}^{AN}$ is given by \eqref{CDF Bm AN final}.


\section*{Appendix~D: Proof of Lemma~\ref{lemma:CDF Ek AN}} \label{Appendix:D}
\renewcommand{\theequation}{D.\arabic{equation}}
\setcounter{equation}{0}
Based on \eqref{SINR e}, the CDF of ${F_{\gamma _{{E_\kappa }}^{AN}}}$ can be expressed as
\begin{align}\label{CDF_Ek_AN}
&{F_{\gamma _{{E_\kappa }}^{AN}}}\left( x \right) = \Pr \left\{ {\mathop {\max }\limits_{e \in {\Phi _e},{d_e} \ge {r_p}} \left\{ {\frac{{{a_\kappa }{P_S}{X_{e,\kappa }}}}{{I_e^{AN} + d_e^\alpha }}} \right\} \le x} \right\}\nonumber\\
&= {E_{{\Phi _e}}}\left\{ {\prod\limits_{e \in {\Phi _e},{d_e} \ge {r_p}} {\int_0^\infty  {{F_{{X_{e,\kappa }}}}\left( {\frac{{\left( {z + d_e^\alpha } \right)x}}{{{a_\kappa }{P_S}}}} \right)} {f_{I_e^{AN}}}\left( z \right)dz} } \right\}.
\end{align}
Following a  procedure similar to that used for obtaining \eqref{gamma XE_CDF 2}, we apply the generating function and switch to polar coordinates. Then \eqref{CDF_Ek_AN} can be expressed as
\begin{align}\label{CDF_Ek_AN 1}
{F_{\gamma _{{E_\kappa }}^{AN}}}\left( x \right) = \exp \left[ { - 2\pi {\lambda _e}\int_{{r_p}}^\infty  {r{e^{ - \frac{x}{{{a_\kappa }{P_S}}}{r^\alpha }}}dr{Q_2}} } \right],
\end{align}
where ${Q_2} = \int_0^\infty  {{e^{ - z\frac{x}{{{a_\kappa }{P_S}}}}}} {f_{I_e^{AN}}}\left( z \right)dz$.
Applying \cite[ Eq. (3.381.9)]{gradshteyn}, we arrive at
\begin{align}\label{CDF_Ek_AN 2}
{F_{\gamma _{{E_\kappa }}^{AN}}}\left( x \right) = \exp \left[ { - \frac{{\mu _{\kappa 1}^{AN}\Gamma \left( {\delta ,\mu _{\kappa 2}^{AN}x} \right)}}{{{x^\delta }}}{Q_2}} \right].
\end{align}

Let us now turn our  attention to solving the integral $Q_2$. Note that all the elements of ${{{\bf{h}}_e}{{\bf{V}}_m}}$ and ${{{\bf{h}}_e}{{\bf{V}}_n}}$ are independent complex Gaussian distributed with a zero mean and unit variance. We introduce the notation ${Y_{e,m}} = {\left\| {{{\bf{h}}_e}{{\bf{V}}_m}} \right\|^2}$ and ${Y_{e,n}} = {\left\| {{{\bf{h}}_e}{{\bf{V}}_n}} \right\|^2}$. As a consequence, both ${Y_{e,m}}$ and ${Y_{e,n}}$  obey the $Gamma\left( {{N_A} - 1,1} \right)$ distribution. Based on the properties of the Gamma distribution, we have ${a_m}\sigma _a^2{Y_{e,m}} \sim Gamma\left( {{N_A} - 1,{a_m}\sigma _a^2} \right),{a_n}\sigma _a^2{Y_{e,n}} \sim Gamma\left( {{N_A} - 1,{a_n}\sigma _a^2} \right)$. Then the sum of these two items $I_e^{AN}$ obeys the generalized integer Gamma (GIG) distribution. According to~\cite{coelho1998generalized}, the PDF of $I_e^{AN}$ is given by
\begin{align}\label{PDF Ek_AN}
{f_{I_e^{AN}}}\left( z \right) = &{\left( { - 1} \right)^{{N_A} - 1}}\prod\limits_{i = 1}^2 {\tau _i^{{N_A} - 1}} \sum\limits_{i = 1}^2 {\sum\limits_{j = 1}^{{N_A} - 1} {} }\nonumber\\
&\frac{{{a_{{N_A} - j,{N_A} - 1}}}}{{\left( {j - 1} \right)!}}{\left( {2{\tau _i} - L} \right)^{j - \left( {2{N_A} - 2} \right)}}{z^{j - 1}}{e^{ - {\tau _i}z}}.
\end{align}
Upon substituting \eqref{PDF Ek_AN} into \eqref{CDF_Ek_AN 2}, as well as applying \cite[ Eq. (3.381.4)]{gradshteyn}, after some further manipulations, we obtain the CDF of ${F_{\gamma _{{E_\kappa }}^{AN}}}$ as
\begin{align}\label{CDF Ek AN final}
{F_{\gamma _{{E_\kappa }}^{AN}}}\left( x \right) = \exp \left[ {\Omega \frac{{\Gamma \left( {\delta ,x\mu _{\kappa 2}^{AN}} \right)}}{{\sum\limits_{p = 0}^j {
 j \choose
 p } {{{{\left( x \right)}^{p + \delta }}{{\left( {{a_\kappa }{P_S}} \right)}^{ - p}}}}\tau _i^{j - p}}}} \right].
\end{align}
Upon setting the derivative of the CDF in \eqref{CDF Ek AN final}, we can obtain \eqref{PDF Ek AN final}.

\bibliographystyle{IEEEtran}
\bibliography{mybib}

\end{document}